\documentclass[aps,amsmath,prd,showpacs,nofootinbib,amssymb,preprint]{revtex4}
\pdfoutput=1
\usepackage{amssymb,graphicx,color,ulem}
\definecolor{cbgreen}{rgb}{0.78,0.08,0.52}

\def\roughly#1{\mathrel{\raise.3ex\hbox
{$#1$\kern-.75em\lower1ex\hbox{$\sim$}}}}

\begin{document}

\title{Complementary Test of the Dark Matter Self-Interaction by Direct and Indirect Detections}

\author{Chian-Shu~Chen$^{1,3}$\footnote{chianshu@gmail.com}, Guey-Lin Lin$^{2}$\footnote{glin@cc.nctu.edu.tw}, and Yen-Hsun Lin$^{2}$\footnote{chris.py99g@g2.nctu.edu.tw}}
  \affiliation{$^{1}$Physics Division, National Center for Theoretical Sciences, Hsinchu 30010, Taiwan\\
$^{2}$Institute of Physics, National Chiao Tung University, Hsinchu 30010, Taiwan\\
$^{3}$Department of Physics, National Tsing Hua University, Hsinchu 30010, Taiwan}

\date{Draft \today}

\begin{abstract}
The halo dark matter (DM) can be captured by the Sun if its final velocity after the collision with a nucleus in the Sun is less than the escape velocity. 
For self-interacting DM (SIDM), we show that the number of DM trapped inside the Sun still provides significant signals even if the DM-nucleon cross section is negligible.
We consider a SIDM model where $U(1)$ gauge symmetry is introduced to account for the DM self-interaction.  Such a model  naturally leads to isospin violating DM-nucleon interaction, although isospin symmetric interaction is still allowed as a special case.    
We show that the detection of neutrino signature from DM annihilation in the Sun can probe those SIDM parameter ranges not reachable 
by direct the detection. 
Those parameter ranges are either the region with a very small $m_{\chi}$ or the region opened up due to isospin violations.    
\end{abstract}

\maketitle

\section{Introduction}\label{sec:introduction}
Single component and collisionless cold dark matter (CCDM), which is treated as the standard dark matter (DM) candidate in $\rm{\Lambda CDM}$ model, accounts for the cosmological data from Cosmic Microwave Background (CMB), Big Bang Nucleosynthesis (BBN), and the large scale structure. However, the above DM properties lead to some controversies between the N-body simulation and astrophysical observations on small scale structures with non-linear DM-dominated systems. A cusp structure is formed by gravitationally attracting a large number of DM in the center regions of galaxies and the zero dissipative dynamics~\cite{Navarro:1996gj}. On the other hand, observations indicate that DM distributes in a much more flat 
profile in the center regions~\cite{Moore:1994yx,Flores:1994gz}. This is called the cusp/core problem~\cite{Walker:2011zu}. There exists other problem with CCDM concerning the size of subhalos, which are the hosts of the satellite galaxies surrounding the Milky Way (MW) halo. The dispersion velocities from these galaxy rotation curves reflect the size of their host subhalos. It is observed that a discrepancy in the subhalo size exists between the CCDM-only simulation and the observations~\cite{Walker:2012td,BoylanKolchin:2011dk}. About $\cal {O}$(10) most massive subhalos generated from the N-body simulation are too massive in the MW halo (with circular velocity larger than 30 km/s) whereas the maximum circular velocities of MW dwarf spheroidals are less than 25 km/s. This is referred to the ``Too-big-to-fail'' problem indicating no satellite galaxy hosted by such massive subhalo is found.  

It is possible that these puzzles may be due to insufficient knowledge of the baryonic processes such as the supernova feedback and photonization~\cite{Efstathiou:1992zz,Efstathiou:1992sy,Mac Low:1998wv,Bullock:2000wn,Hopkins:2011rj,Governato:2012fa,Brooks:2012vi}. On the other hand, these conflicts may also 
hint nontrivial features of DM such as self-interacting DM  (SIDM)~\cite{Spergel:1999mh}. It was then noted that the SIDM with a constant cross section cannot account for observed    
ellipticities in clusters~\cite{Yoshida:2000uw,MiraldaEscude:2000qt} and the survivability of subhalos~\cite{Gnedin:2000ea}. In the very short mean free path limit, it produces even more cuspy profiles~\cite{Mo:2000pu,Firmani:2000ce,Moore:2000fp,Yoshida:2000bx}. However, some recent simulations have shown that the SIDM velocity-independent cross sections in the range $0.1 \ {\rm cm^2/g}\le \sigma_{T}/m_{\chi} \le 10$~$\rm{cm^2/g}$ ($m_{\chi}$ denoting the DM mass) can resolve the cusp/core and Too-big-to-fail problems on dwarf scales~\cite{Vogelsberger:2012ku,Zavala:2012us,Rocha:2012jg,Peter:2012jh}, although it was pointed out in Ref.~\cite{Zavala:2012us} that $\sigma_{T}/m_{\chi} \sim 0.1$~$\rm{cm^2/g}$ is too small to account for the population of massive subhalos, and the study using bullet cluster 1E 0657-56 constrains $\sigma_{T}/m_{\chi}$ to be less than 1.25~$\rm{cm^2/g}$~\cite{Randall:2007ph}. Furthermore, other investigations of SIDM with characteristic velocity-dependent cross sections provide a broader cross section range, $0.1 \ {\rm cm^2/g}\le \sigma_{T}/m_{\chi} \le 50$~$\rm{cm^2/g}$, for alleviating the above-mentioned puzzles~\cite{Feng:2009hw,Buckley:2009in,Aarssen:2012fx,Tulin:2012wi,DelNobile:2015uua}\footnote{The thorough study on cosmology and structure formation in the context of mirror photon is investigated in Ref.~\cite{Foot:2014uba}. Some early attempts to study the astrophysical effects of mirror matters can also be found, for example, in Ref.~\cite{S.I.Blinnikov}.}. From particle physics point of view, DM exchanging light mediators would generate an attractive self-interaction, which enhances the annihilation cross section by Sommerfield effect and resonance scattering~\cite{Cirelli:2008pk,ArkaniHamed:2008qn}. The mediator $\phi$ often plays as the messenger between the visible matter and dark matter. This mediator can be scalar, pseudoscalar, vector, or axial-vector particles~\cite{Bellazzini:2013foa}. The scalar and vector interactions with nucleon only generate spin-independent (SI) cross section (in the non-relativistic limit), while spin-dependent cross section can be generated via exchanging pseudoscalar or axial-vector particles. The relatively large SIDM cross section requires a MeV scale $\phi$ satisfying~\cite{Feng:2009hw} 
\begin{equation}
\sigma_{\chi\chi}\approx7.6\times10^{-24}\,{\rm cm}^{2}\,\left(\frac{\alpha_{\chi}}{0.01}\right)^{2}\left(\frac{m_{\phi}}{30\,{\rm MeV}}\right)^{-4}\left(\frac{m_{\chi}}{{\rm GeV}}\right)^{2}~,\label{eq:sig_XX}
\end{equation} 
where $\alpha_{\chi}$ is the dark fine structure constant representing the coupling strength between DM $\chi$ and the mediator $\phi$. 
With $m_{\phi}$ around the MeV scale, it is possible for $\phi$ to decay into light standard model particles. In such a case, 
the lifetime of $\phi$ should be less than 1 sec to satisfy the BBN constraint.\footnote{If DM is kinematically decoupled from the dark radiation at late time, it would leave signals in CMB~\cite{CyrRacine:2012fz}.} Therefore, the light mediator $\phi$ can produce both direct and indirect DM signatures. The former signature is generated when DM scatters with the nuclei by exchanging $\phi$. The latter signature is generated by the annihilation of DMs into a pair of $\phi$, which subsequently decay into SM particles. The mediator $\phi$ can be a hidden $U(1)$ gauge boson or scalar, which mixes with photon, $Z$ boson or Higgs boson to bridge between DM and the visible sectors. It should be noted that Eq.~(\ref{eq:sig_XX}) is valid only at the perturbative regime ($\alpha_{\chi}m_{\chi}/m_{\phi} \le 1$), while the effects of dark force on the halo structure for the full range of parameters were investigated in Ref.~\cite{Tulin:2012wi}. The characteristic velocity-dependent cross section at different scales for resolving all the conflicts can be realized. Furthermore, DM direct detection experiments for probing such SIDM scenarios were proposed in Ref.~\cite{Kaplinghat:2013yxa}~(see also, for example, Ref.~\cite{Laha:2013gva} for the asymmetric DM scenario). The predicted direct detection cross section is within the reach of next-generation experiments such as XENON1T~\cite{Aprile:2012zx} and SuperCDMS~\cite{Brink:2012zza}.  

In this paper, we shall consider symmetric fermionic DM and show that the annihilation signatures from the trapped DMs inside the Sun can provide a complementary test to SIDM scenarios.
It is well known that the DM-nucleon scattering cross section relevant to the indirect DM signature from the Sun is identical to the  
cross section relevant to DM direct detection experiments. For the former case, such a cross section leads to the DM capture if
the velocity of the final-state DM is less than the escape velocity of the Sun. As the number of captured DMs increases to a 
significant level,  the rate of DM annihilations into SM particles could become detectable~\cite{Silk:1985ax,Srednicki:1986vj,Spergel:1984re,
Press:1985ug,Griest:1986yu,Gould:1987ju,Gould:1991hx,
Zentner:2009is,Albuquerque:2013xna,Chen:2014oaa}. 
The total number of the captured DMs is determined by the above DM-nucleon scattering cross section and the DM mass. However, it has been shown in a model-independent approach that the DM accumulation time can be shortened in the Sun and the total  number of trapped DM can also be increased if SIDM is considered~\cite{Zentner:2009is,Albuquerque:2013xna,Chen:2014oaa}. In Ref.~\cite{Albuquerque:2013xna}, the constraint $\sigma_{T} < \cal{O}$$(10^{-22}) \ \rm{cm^2}$ is obtained for the $\chi\chi \rightarrow W^{+}W^{-}/\tau^{+}\tau^{-}$ annihilation mode using IceCube 79 string data~\cite{Aartsen:2012kia} which probes the mass range $m_{\chi} >$ 20 GeV. The low-mass DM region to be probed by IceCube-PINGU is considered in Ref.~\cite{Chen:2014oaa}.  In this work we shall present a realization of the above model-independent results with a concrete SIDM model. We work out the SIDM model prediction on the DM-induced neutrino flux from the Sun and the sensitivities of future neutrino telescopes to such a signal and the related model parameters. We compare these sensitivities to constraints set by current direct detection experiments.
We will show that the searches for direct and indirect DM signals give complementary tests to the parameter space of SIDM model.     
The complementarity of direct and indirect searches arises from two reasons. First, the indirect search can look for very light DM which is not yet probed by
direct detections.  Second, the direct detection sensitivity could be significantly worsen by isospin violation in DM-nucleon couplings.  On the other hand, 
the DM-induced neutrino flux from the Sun can be insensitive to DM-nucleus scattering cross section so long as DM self-interaction is 
strong enough~\cite{Zentner:2009is,Chen:2014oaa}.

Our paper is organized as follows. In Section II we briefly introduce the SIDM models where the light mediator is either a gauge boson or a  scalar particle. The constraints on these models are summarized. Section III shows that the DM accumulation inside the Sun can be significantly enhanced by DM self-interaction. In Section IV, we  demonstrate that  IceCube/PINGU experiment can probe SIDM parameters in a way complementary to the direct detection experiment. We present our conclusions in Section V.

\section{Scattering via light mediator exchange}\label{sec:mediator}
Two simplest ways to construct the SIDM model with the feature of Sommerfield enhancement are to assume a light vector boson or a light scalar boson as the dark force mediator\footnote{In this paper, we analyze the case of velocity-independent cross sections.}. For the vector mediator case, DM carries a charge $e_{D}$ under a hidden Abelian  $U(1)_{\chi}$ gauge symmetry and $\phi^{\mu}$ is the corresponding gauge boson. The hidden $U(1)_{\chi}$ gauge boson has no ordinary SM interaction except the one induced by kinetic mixing with the SM photon~\cite{Holdom:1985ag}. Since we expect the mass of dark mediator $\phi$ to be around MeV scale, there must exist certain mechanism to generate a massive $U(1)_{\chi}$ gauge boson. It can be certain spontaneous symmetry breaking pattern in the  dark sector or based upon the Stueckelberg mechanism~\cite{Stueckelberg:1900zz,Stueckelberg:1938zz} to ensure the UV completion of the theory. Here we take the phenomenological approach and neglect the details of model construction. The $\phi \gamma$ and $\phi Z$ mixings lead to the following couplings between $\phi^{\mu}$ and SM currents:
\begin{eqnarray}
\mathcal{L}_{\rm mixing/vector} = \left( \epsilon_{\gamma}eJ^{\mu}_{\rm em} + \epsilon_{\rm Z}\frac{g_{2}}{c_W}J^{\mu}_{\rm NC}\right) \phi_{\mu} , 
\end{eqnarray}
with 
\begin{eqnarray}
J^{\mu}_{\rm em} = \sum_{f}Q_{f}\bar{f}\gamma^{\mu}f   \quad {\rm and} \quad 
J^{\mu}_{\rm NC} = \sum_{f}\bar{f}\gamma^{\mu}\left[ I_{3f}\left(\frac{1 - \gamma_{5}}{2}\right) - Q_{f}s^2_{W}\right] f   
\end{eqnarray}
representing electromagnetic current and weak neutral current respectively. Here $f$ is the SM fermion and $Q_{f}$ refers to its electric charge, $g_{2}$ is the $SU(2)_{L}$ gauge coupling, and $c_{W}(s_{W}) = \cos{\theta_{W}}(\sin{\theta_{W}})$. The parameters $\varepsilon_{\gamma}$ and $\varepsilon_{Z}$ are originated from kinetic and mass mixings between $\phi^{\mu}$ and gauge bosons such as
\begin{eqnarray}
\mathcal{L}_{{\rm mixing},\,U(1)} = \frac{\varepsilon_{\gamma}}{2}\phi_{\mu\nu}F^{\mu\nu} + \varepsilon_{Z} m_{Z}^2\phi_{\mu}Z^{\mu}.
\end{eqnarray}
With DM $\chi$ a Dirac fermion, the dark force is given by 
\begin{eqnarray}
\mathcal{L}_{\rm DM_{\phi_{vector}}} = e_{D}\bar{\chi}\gamma^{\mu}\chi\phi_{\mu}.  
 \end{eqnarray}  

For the case of scalar $\phi$, it is possible that $\phi$ can mix with Higgs boson. The simplest scenario is that $\phi$ is a real singlet scalar. The relevant terms of the Lagrangian are then given by
 \begin{eqnarray}
 \mathcal{L}_{\rm DM_{\phi_{scalar}}} \supset g_{s}\bar{\chi}\chi\phi + a\phi|H|^2 + b\phi^2|H|^2 , 
 \end{eqnarray}
where $g_{s}$, $a$, and $b$ are coupling constants and $H$ is the SM Higgs doublet. As the Higgs boson develops the vacuum expectation value $v \approx 246$~GeV, the mass matrix between $\phi$ and $H$ can mix via the $a$ term. We can define the mixing parameter $\varepsilon_{h} \sim av/m^2_{h}$,  thus the effective Lagrangian for the coupling of scalar $\phi$ to the SM fermions is given by  
\begin{eqnarray}
\mathcal{L}_{\rm mixing/scalar} = \sum_{f}-\varepsilon_{h}\frac{m_{f}}{v}\bar{f}f .
 \end{eqnarray} 
\subsection{Scattering with vector mediator\label{sub:XX-scat}}
\begin{figure}
\begin{centering}
	\includegraphics[width=0.2\textwidth]{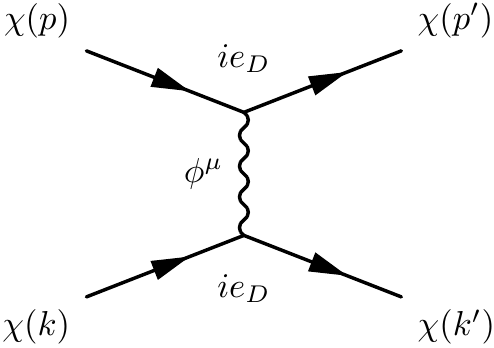}\quad{}\includegraphics[width=0.2\textwidth]{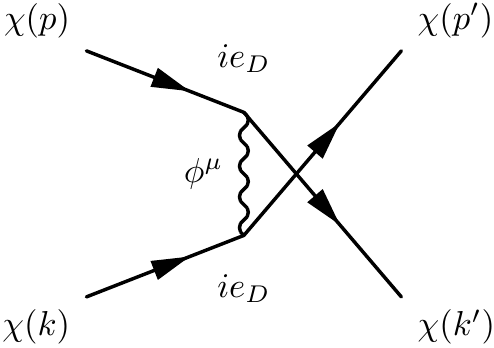}
	\par\end{centering}

\begin{centering}
	\includegraphics[width=0.2\textwidth]{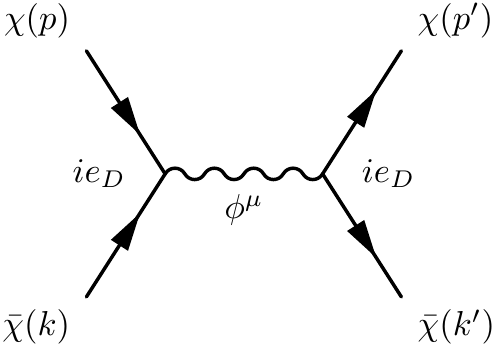}\quad{}\includegraphics[width=0.2\textwidth]{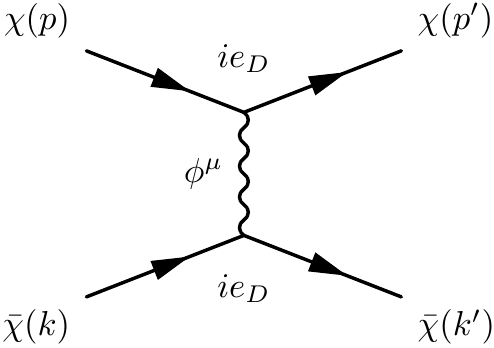}
	\par\end{centering}

\protect\caption{\label{fig:chi-chi_scat}The Feynman diagrams contributing to DM scatterings. The upper panel
represents the $t$ and $u$ channel diagrams of $\chi\chi$ scattering. The lower one
represents the $s$ and $t$ channel diagrams of $\chi\bar{\chi}$ scattering. The $\chi\chi$
cross section vanishes in the non-relativistic limit while $\chi\bar{\chi}$ scattering is dominated by the $t$ channel diagram. The behavior of 
$\bar{\chi}\bar{\chi}$ scattering is the same as that of $\chi\chi$ scattering. See the main text for details.}
\end{figure}

The Feynman diagrams for scattering among two Dirac fermionic (anti-) DMs via the dark mediator $\phi$ 
are depicted in Fig.~\ref{fig:chi-chi_scat}. Due to Fermi statistics, the $u$ and $t$ channel amplitudes of $\chi\chi$ scattering carry a relative minus sign such that two amplitudes cancel each other
in the non-relativistic limit. The same cancellation occurs in $\bar{\chi}\bar{\chi}$ scattering. Hence only $\chi\bar{\chi}$ scattering cross section is nonvanishing in the non-relativistic limit.
Furthermore, 
the $s$ channel amplitude of  $\chi\bar{\chi}$ scattering is suppressed so long as $m_{\chi}\gg m_{\phi}$. Hence we consider only $t$ channel amplitude for $\chi\bar{\chi}$ scattering.
The perturbative calculation gives the cross section
\begin{equation}
\sigma_{\chi\bar{\chi}}\approx4\pi\alpha_{\chi}^{2}\frac{m_{\chi}^{2}}{m_{\phi}^{4}}~,\label{eq:sig_X-barX}
\end{equation}
where $\alpha_{\chi} \equiv e_{D}^{2}/(4\pi)$ is the fine structure constant
in the hidden $U(1)$ sector. 
Since only $\sigma_{\chi\bar{\chi}}$ is not suppressed for Dirac DM, the parameterization in Eq.~(\ref{eq:sig_XX}) should be understood as the one for  $\sigma_{\chi\bar{\chi}}$.
In fact, for simplifying the notations, we henceforth denote the $\chi\bar{\chi}$ scattering cross section as $\sigma_{\chi\chi}$.

\begin{figure}
\begin{centering}
	\includegraphics[width=0.2\textwidth]{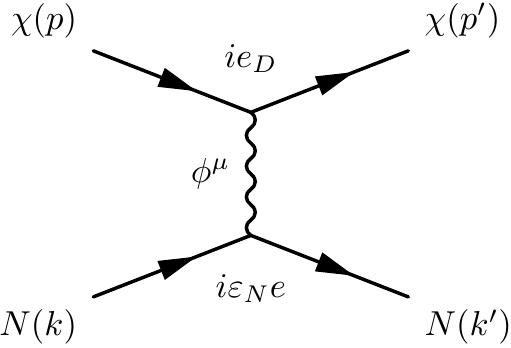}
	\par\end{centering}

\protect\caption{\label{fig:chi-p_scat}The Feynman diagram contributing to $\chi $-$N$ scattering. 
One simply replaces $\chi$ by $\bar{\chi}$ in the diagram for $\bar{\chi}$-$N$ scattering.}
\end{figure}

The DM-nucleon scattering is shown in Fig.~\ref{fig:chi-p_scat}
where $N$ stands for either proton or neutron. The parameter $\varepsilon_{N}$ is the strength of $\phi_{\mu}$-nucleon coupling 
in the unit of electric charge $e$. The SI cross section between DM and any nucleus
with mass number $A$ and proton number $Z$ is then given by 
\begin{equation}
\sigma_{\chi A}\approx\frac{16\pi\alpha_{\chi}\alpha_{{\rm em}}}{m_{\phi}^{4}}[\varepsilon_{p}Z+\varepsilon_{n}(A-Z)]^2\mu_{\chi A}^{2}= \frac{16\pi\alpha_{\chi}\alpha_{{\rm em}}}{m_{\phi}^{4}}\varepsilon_p^2[Z+\eta(A-Z)]^2\mu_{\chi A}^{2} ,\label{eq:sig_XA}
\end{equation}
with $\alpha_{\rm em}$ the fine structure
constant, $\mu_{\chi A}\equiv m_{\chi}m_{A}/(m_{\chi}+m_{A})$ the
reduced mass for DM-nucleus system and $\eta\equiv \varepsilon_n/\varepsilon_p$ is the isospin violation parameter. The cross section for $\bar{\chi}$ is identical such that $\sigma_{\chi A}=\sigma_{\bar{\chi}A}$. The effective $\phi_{\mu}$-nucleon couplings, $\varepsilon_{p}$ and $\varepsilon_{n}$, can be expressed in terms of $\varepsilon_{\gamma}$ and $\varepsilon_{Z}$ such that~\cite{Kaplinghat:2013yxa}
\begin{subequations}
\begin{align}
\varepsilon_{p} & =\varepsilon_{\gamma}+\frac{\varepsilon_{Z}}{4s{}_{W}c{}_{W}}(1-4s_{W}^{2})\approx\varepsilon_{\gamma}+0.05\varepsilon_{Z}~,\label{eq:e_p}\\
\varepsilon_{n} & =-\frac{\varepsilon_{Z}}{4s_{W}c_{W}}\approx-0.6\varepsilon_{Z}~.\label{eq:e_n}
\end{align}
\end{subequations}
Thus, we have SI cross sections
\begin{equation}
\sigma^{\rm SI}_{\chi p}\approx1.5\times10^{-24}\,{\rm cm}^{2}\,\varepsilon_{\gamma}^{2}\left(\frac{\alpha_{\chi}}{0.01}\right)\left(\frac{m_{\phi}}{30\,{\rm MeV}}\right)^{-4}\label{eq:sig_XP}
\end{equation}
for DM-proton scattering and
\begin{equation}
\sigma^{\rm SI}_{\chi n}\approx5\times10^{-25}\,{\rm cm}^{2}\,\varepsilon_{Z}^{2}\left(\frac{\alpha_{\chi}}{0.01}\right)\left(\frac{m_{\phi}}{30\,{\rm MeV}}\right)^{-4}\label{eq:sig_XN}
\end{equation}
for DM-neutron scattering.

\subsection{Scattering with scalar mediator}

The Feynman diagrams for DM-DM scattering with scalar exchange are
identical to those depicted in Fig.~\ref{fig:chi-chi_scat}. Thus we only replace $\alpha_{\chi}$ by $\alpha_s = g^2_{s}/4\pi$, and the mixing parameter $\varepsilon_h$ via the Higgs boson are the same for both  proton and neutron. Roughly one has 
\begin{equation}
\varepsilon_{p,n}\approx3\times10^{-3}\varepsilon_{h} .
\end{equation}
The general DM-nucleon cross section thus obeys the isospin symmetry with
\begin{equation}
\sigma_{\chi N}\approx2\times10^{-29}\,{\rm cm}^{2}\,\varepsilon_{h}^{2}\left(\frac{\alpha_{s}}{0.01}\right)\left(\frac{m_{\phi}}{30\,{\rm MeV}}\right)^{-4}.
\end{equation}
However, to have $\phi$ decay before BBN, one requires $\varepsilon_{h}\gtrsim10^{-5}$. With such 
a lower bound for $\varepsilon_h$, the resulting  $\sigma_{\chi p}$ is so large that it is 
excluded by the current direct search for $m_{\chi}>10\,{\rm GeV}$ \cite{Kaplinghat:2013yxa}. In
this work, we shall not continue discussing  the scalar case, although some other scalar-exchange models might still be viable. It is important to note that the vector mediator case already contains many features of SIDM models. In addition, we are interested in searching for neutrino signals. Such type of signals disfavors the scalar mediator since the decay branching ratios of $\phi$ to neutrinos are generally negligible, unless $m_{\phi} < 1$~MeV. We note that the spin-dependent DM-nucleon cross section is suppressed by ${\cal O} (1/m_{\chi})$ for the case of vector mediator.  On the other hand, such suppression does not occur if $\phi$ is an axial-vector particle.
In this paper, we shall only focus on spin-independent cross section. Hence we simplify $\sigma_{\chi p}^{\rm SI}$ as $\sigma_{\chi p}$ from now on.

\section{DM accumulation in the Sun}

\subsection{The evolution equation}
Since we consider the scenario that $\chi$ and $\bar{\chi}$ are equally populated, 
the halo DM number density near the solar system can then be written as $\rho_{0}=\rho_{\chi}+\rho_{\bar{\chi}}=0.3\,{\rm GeV}\,{\rm cm}^{-3}$
with $\rho_{\chi}=\rho_{\bar{\chi}}=0.15\,{\rm GeV}\,{\rm cm}^{-3}$. 
The DM in the Sun consists of two species, the DM and anti-DM with the numbers $N_{\chi}$ and $N_{\bar{\chi}}$ respectively. The time evolutions of $N_{\chi}$ and $N_{\bar{\chi}}$ 
are given by 
\begin{subequations}
\begin{align}
\frac{dN_{\chi}}{dt} & =C_{c}-C_{e}N_{\chi}+C_{s}N_{\bar{\chi}}-(C_{a}+C_{se})N_{\chi}N_{\bar{\chi}},\label{eq:n_chi}\\
\frac{dN_{\bar{\chi}}}{dt} & =C_{c}-C_{e}N_{\bar{\chi}}+C_{s}N_{\chi}-(C_{a}+C_{se})N_{\bar{\chi}}N_{\chi},\label{eq:n_chi_bar}
\end{align}
\end{subequations}
with $C_{c}$ the capture rate, $C_{e}$
the evaporation rate, $C_{s}$ the capture rate due to self-interaction, $C_{se}$ the self-interaction
induced evaporation rate, and $C_{a}$ the annihilation rate. They are taken
to be time-independent and have been fully discussed in Refs.~\cite{Spergel:1984re,
Press:1985ug,Griest:1986yu,Gould:1987ju,Gould:1991hx,
Zentner:2009is,Chen:2014oaa} and references therein.

\begin{figure}
	\begin{centering}
		\includegraphics[width=0.43\textwidth]{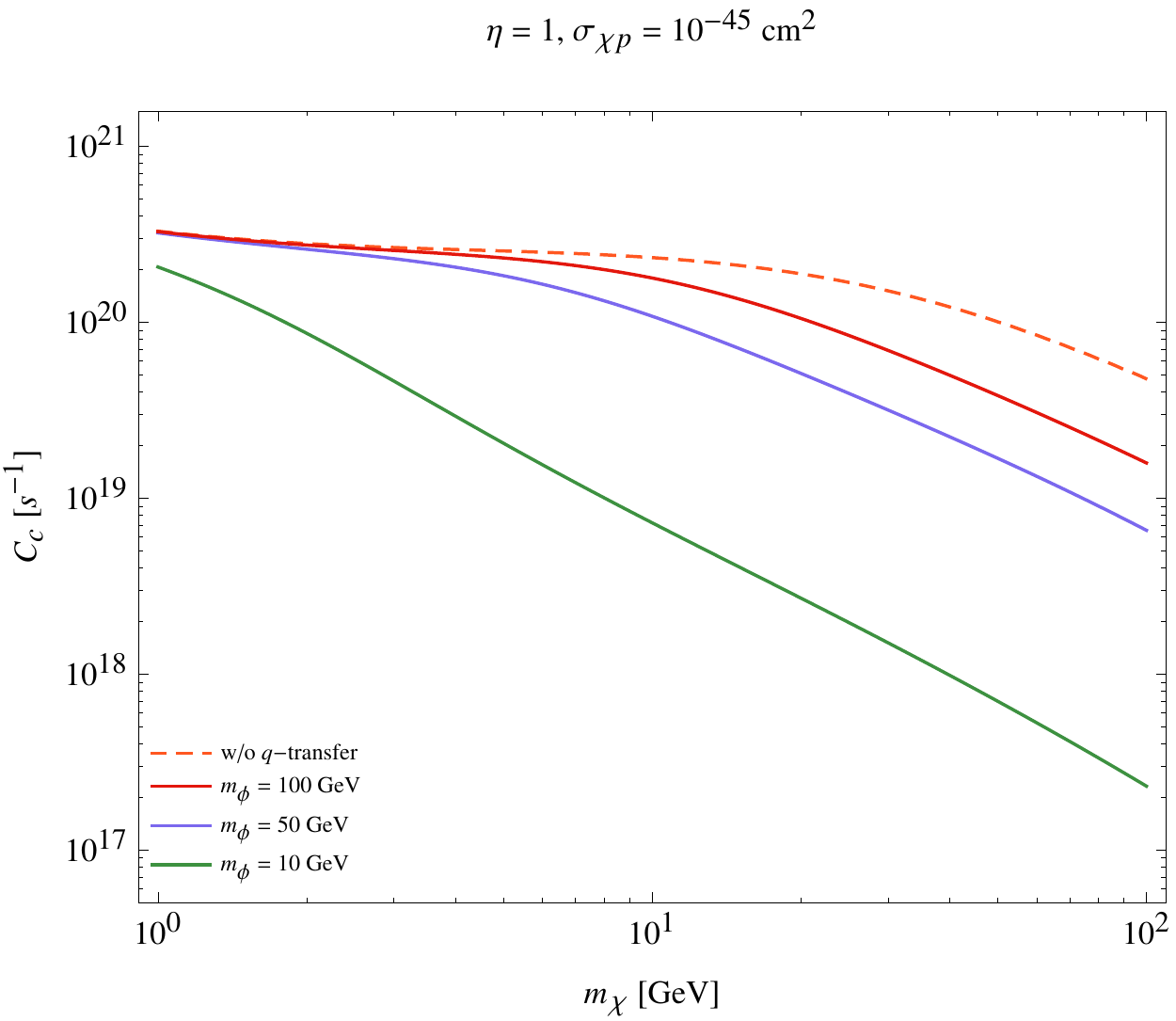}\quad{}\includegraphics[width=0.43\textwidth]{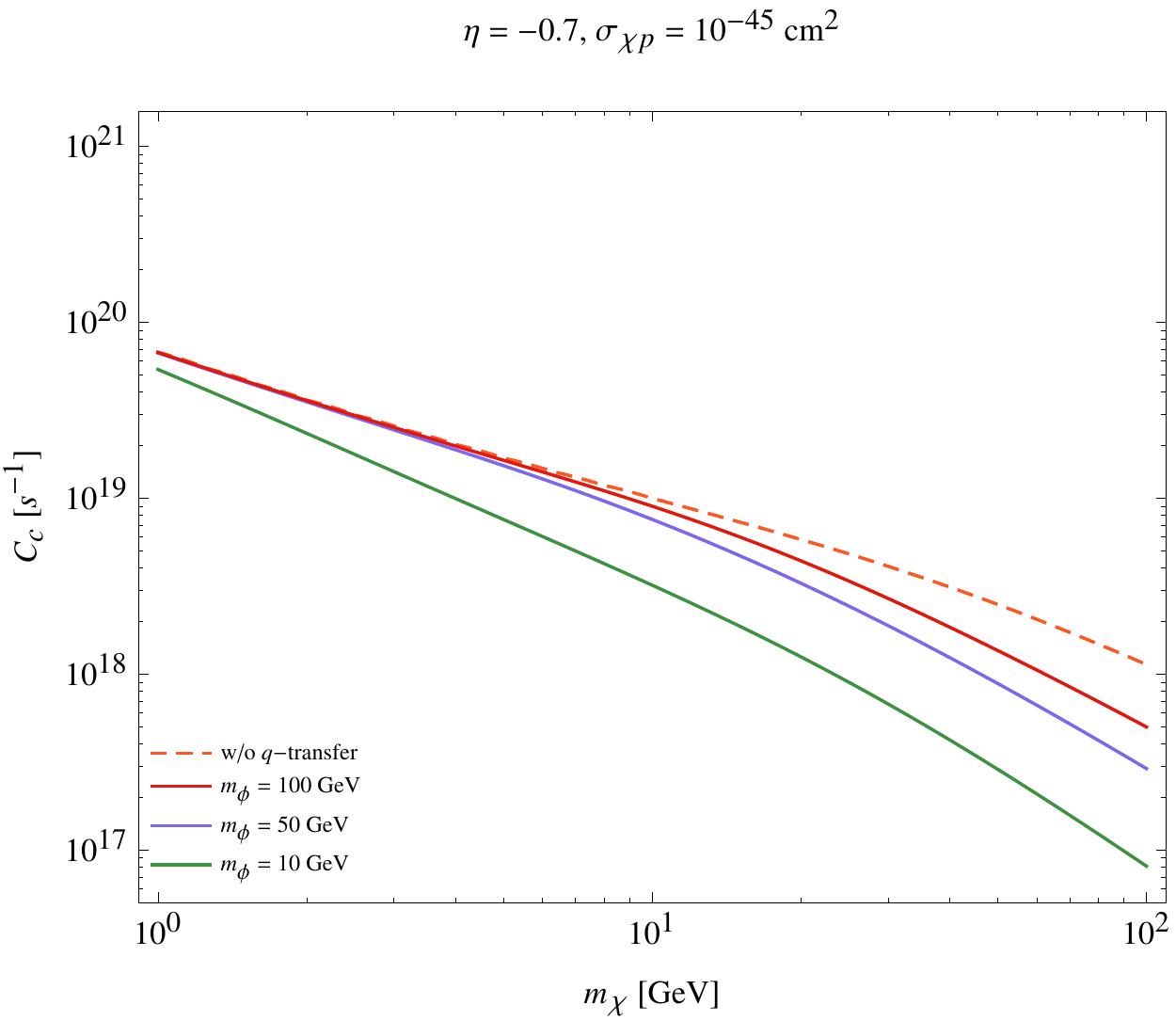}
		\par\end{centering}
	\protect\caption{\label{fig:CaptureRate}The capture rates with (solids) and without momentum transfer (dashed). The cross section $\sigma_{\chi p}^0=10^{-45}~\rm cm^2$ for all results. Left and right are corresponding to $\eta=1$ and $-0.7$.}
\end{figure}

The capture rates $C_c$ and $C_s$ depend on DM-nucleus scattering cross section $\sigma_{\chi A}$
and DM-DM self-interaction cross section $\sigma_{\chi\chi}$, respectively.
For a MeV range $m_{\phi}$, it has been pointed out that~\cite{Kaplinghat:2013yxa,Fornengo:2011sz} 
$\sigma_{\chi A}$ is sensitive to the momentum transfer $\boldsymbol{q}$ flowing into the 
$\phi$ propagator shown in Fig.~\ref{fig:chi-p_scat}. We thus have 
\begin{equation}\label{eq:sigma_q}
\sigma_{\chi A}(\boldsymbol{q}^2) = \frac{m_\phi^4}{(m_\phi^2+\boldsymbol{q}^2)^2}\sigma_{\chi A}^0
\end{equation}
where the magnitude of momentum transfer is given by $\boldsymbol{q}^2=2m_A E_R$ with $E_R$ the 
nucleus recoil energy, and $\sigma_{\chi A}^0\equiv \sigma_{\chi A}(\boldsymbol{q}^2=0)$ is the cross section with zero momentum transfer. It is clearly seen that $\sigma_{\chi A}(\boldsymbol{q}^2)$ is suppressed compared to
$\sigma_{\chi A}^0$. For estimating this suppression, we take $E_R=2\mu_{\chi A}^2v_\chi^2/m_A$ with $\mu_{\chi A}$ the reduced mass for the DM-nucleus system.  This $E_R$
corresponds to an $180^{\circ}$ DM recoil angle after the collision in the center of momentum frame. We further take $v_\chi \approx 270~\rm km/\rm s$, which is the DM velocity dispersion in the halo. 
Apparently, $\boldsymbol{q}^2$ depends on both $m_A$ and $m_{\chi}$. Thus, the capture rate with the momentum transfer suppression can be expressed as
\begin{equation}\label{eq:CcQ}
C_c \propto \left(\frac{\rho_\chi}{0.15~\rm GeV/\rm cm^{3}}\right)\left(\frac{\rm GeV}{m_\chi}\right)\left(\frac{270~\rm km/\rm s}{v_\chi}\right)\sum_A F_A(m_\chi,\eta) \sigma_{\chi A}^0\frac{m_\phi^4}{(m_\phi^2+\boldsymbol{q_A}^2)^2},
\end{equation}
where $F_A(m_\chi,\eta)$ is the product of various factors relevant to the chemical element $A$ in the Sun, including the mass fraction, chemical element distribution, kinematic suppression, form factor, reduced mass and isospin violation effect. 
In Fig.~\ref{fig:CaptureRate}, we compare the effect of momentum transfer suppression on $C_c$ for 
different $m_\phi$ with $\sigma_{\chi p}^0$ fixed at $10^{-45}$ cm$^2$. One can see that the momentum transfer suppression 
is more severe for larger $m_{\chi}$ since $\boldsymbol{q_A}^2$ increases with $m_{\chi}$. 

The above momentum transfer suppression also occurs in $C_s$. The suppression factor can be taken from Eqs.~(\ref{eq:sigma_q}) and (\ref{eq:CcQ}). Thus
\begin{equation}
C_s(\boldsymbol{q}^2)=C_s(\boldsymbol{q}^2=0)\frac{m_\phi^4}{(m_\phi^2+\boldsymbol{q}^2)^2}.
\end{equation}
For DM self-capture, we have $\boldsymbol{q}=m_\chi (\boldsymbol{v_i}-\boldsymbol{v_f})$ where $\boldsymbol{v_i}$ is the DM velocity before the scattering with the magnitude of the velocity given by $v_i=\sqrt{v_\chi^2+v_{\rm esc}(r)^2}$, while $\boldsymbol{v_f}$ is the DM velocity after the scattering with $v_f=v_{\rm esc}(r)/\sqrt{2}$ on average~\cite{Chen:2015poa}. We take the approximation that $\boldsymbol{v_f}$ is parallel to $\boldsymbol{v_i}$
since the speed halo DM is in general much greater than that of the trapped DM.

Having discussed the properties of $C_c$ and $C_s$, we return to Eqs. (\ref{eq:n_chi}) and (\ref{eq:n_chi_bar}).  
Since we are interested in the symmetric DM, the condition $N_\chi=N_{\bar{\chi}}$ holds.
Thus, Eqs.~(\ref{eq:n_chi}) and (\ref{eq:n_chi_bar}) can be
simplified into one,
\begin{equation}
\frac{dN_{\chi}}{dt}=C_{c}-C_{e}N_{\chi}+C_{s}N_{\chi}-(C_{a}+C_{se})N_{\chi}^{2},\label{eq:evo_eq}
\end{equation}
with the solution 
\begin{equation}
N_{\chi}(t)=\frac{C_{c}\tanh(t/\tau_{A})}{\tau_{A}^{-1}-(C_{s}-C_{e})\tanh(t/\tau_{A})/2} ,
\end{equation}
where
\begin{equation}
\tau_{A}\equiv \frac{1}{\sqrt{C_{c}(C_{a}+C_{se})+(C_{s}-C_{e})^{2}/4}}\label{eq:eq_time}
\end{equation}
is the time-scale for the DM number in the Sun to reach the equilibrium, i.e., $dN_\chi/dt=0$.
The DM number reaches the equilibrium when $\tanh(t/\tau_{A})\sim1$.

The DM annihilation rate in the Sun is given by
\begin{equation}
\Gamma_{A}=C_{a}N_{\chi}N_{\bar{\chi}}=C_a N_\chi^2.
\end{equation}
By setting $C_{s}=C_{se}=0$, we
recover the results in Refs.~\cite{Spergel:1984re,
Press:1985ug,Griest:1986yu,Gould:1987ju,Gould:1991hx} for the absence of DM self-interaction. By setting 
$C_{e}=C_{se}=0$, we recover the result in Ref.~\cite{Zentner:2009is}, which
includes the DM self-interaction while neglects the DM evaporation.

\subsection{Numerical results}\label{sub:numerical}
\begin{figure}
\begin{centering}
\includegraphics[width=0.43\textwidth]{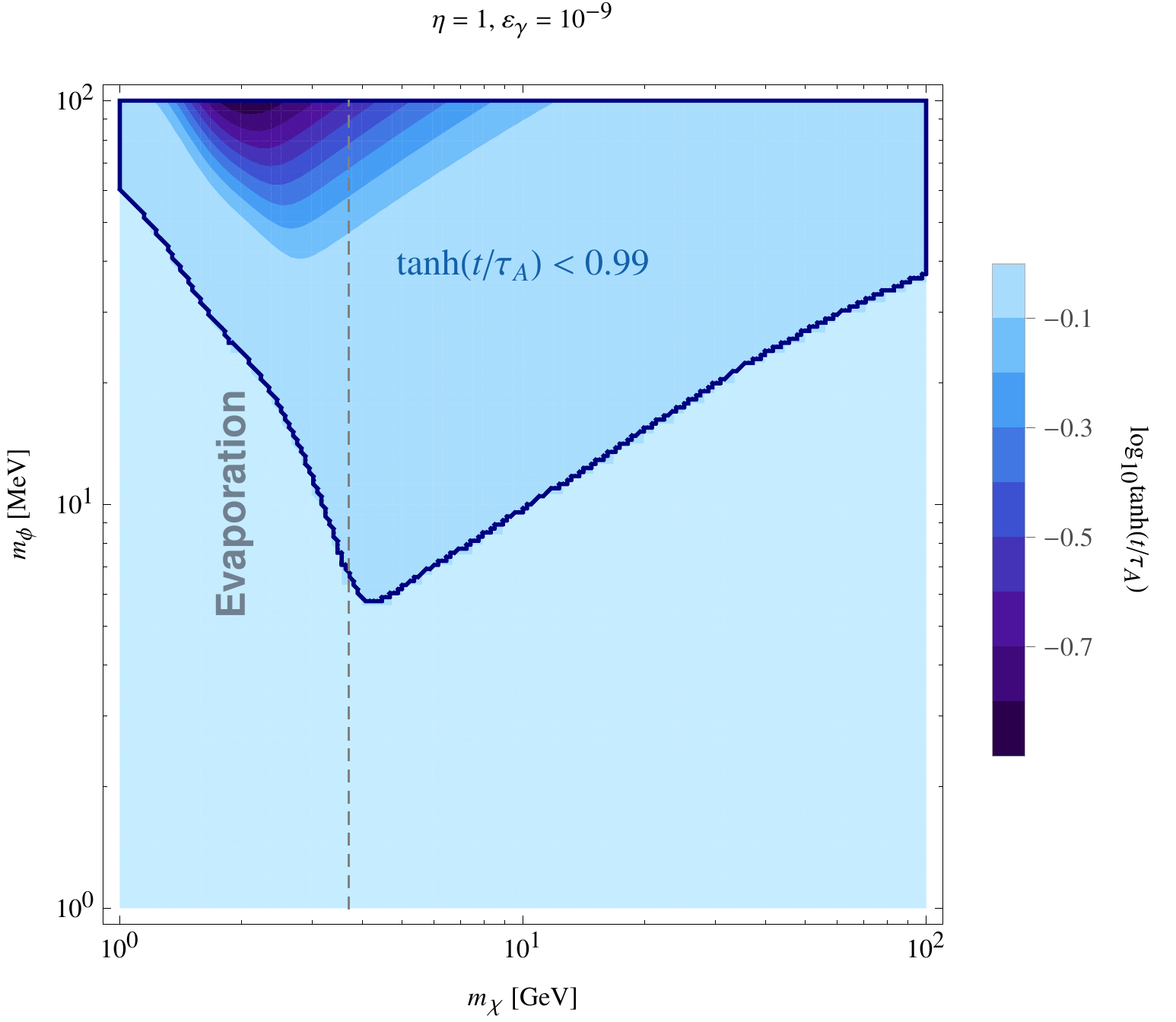}\quad{}\includegraphics[width=0.43\textwidth]{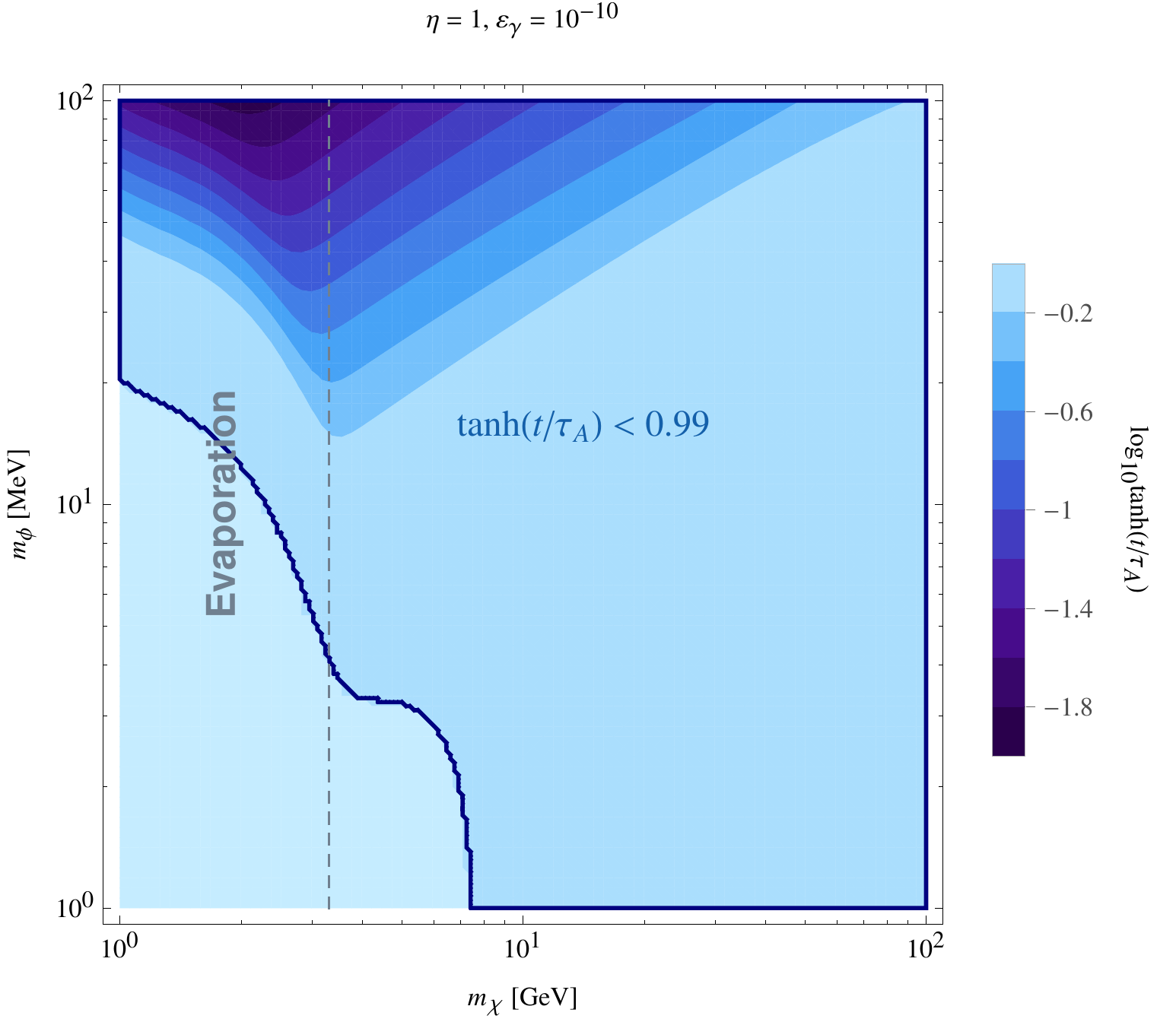}
\par\end{centering}

\begin{centering}
\includegraphics[width=0.43\textwidth]{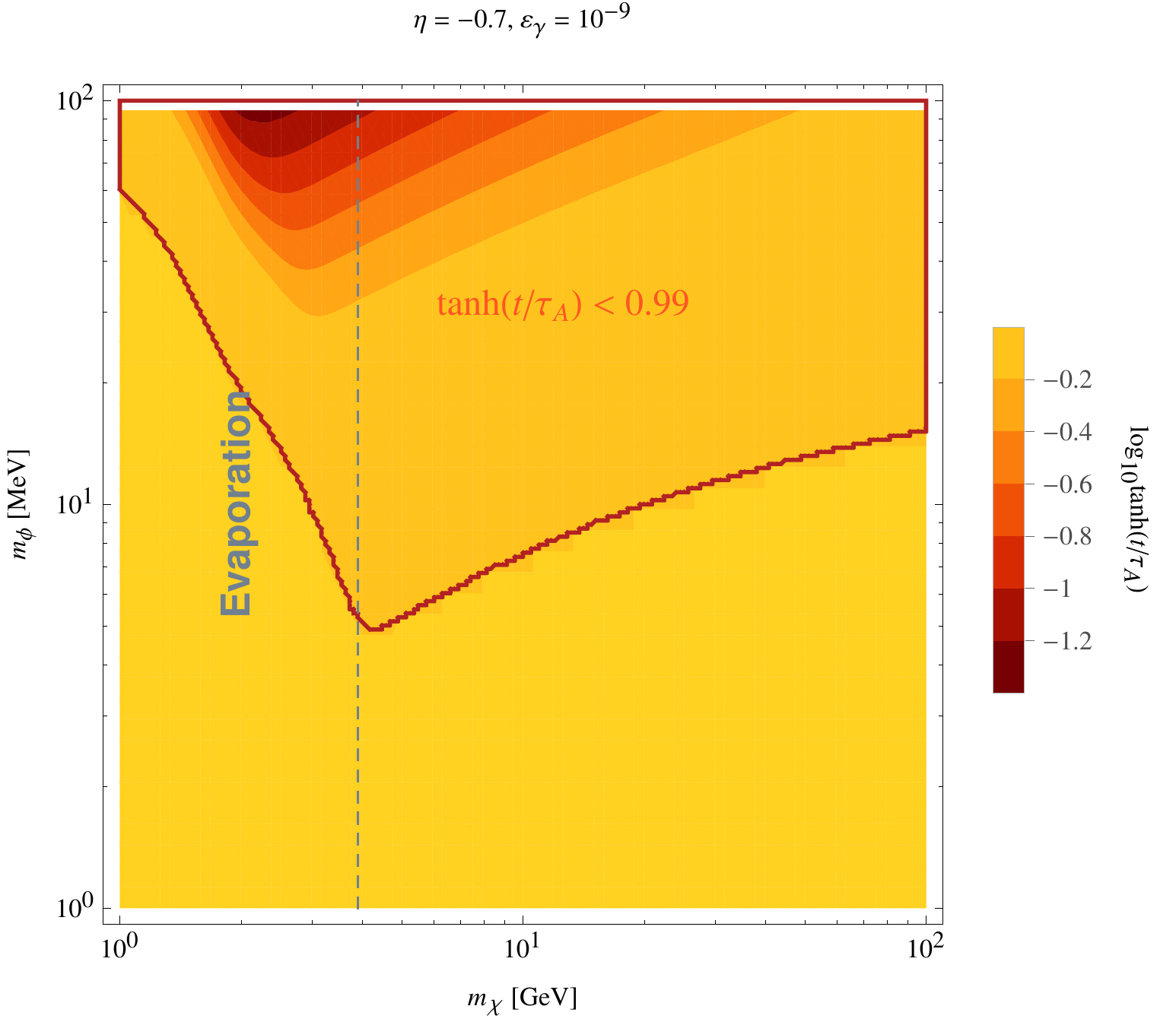}\quad{}\includegraphics[width=0.43\textwidth]{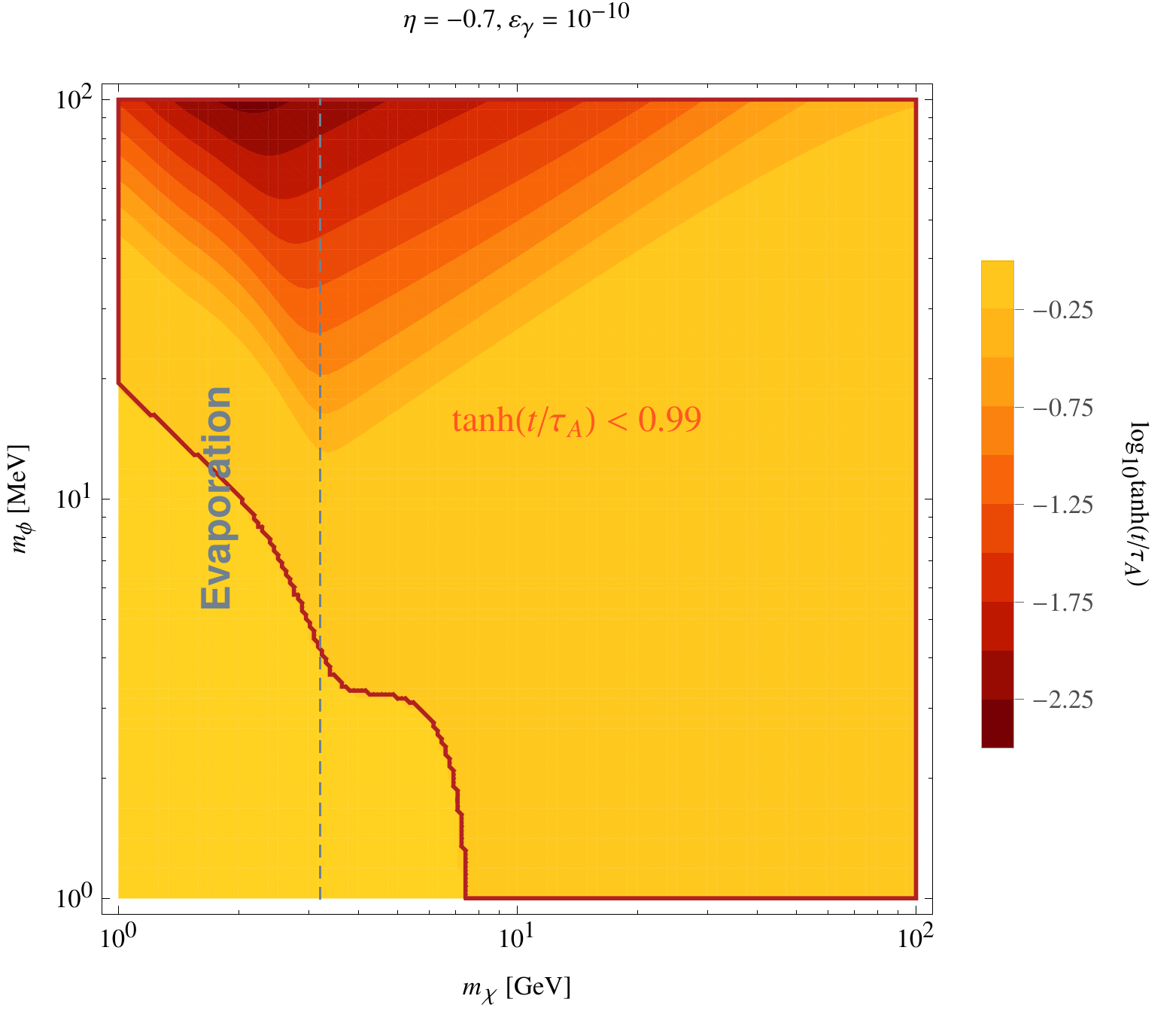}
\par\end{centering}
\protect\caption{\label{fig:eq_time}The dependencies of $\tanh(t_{\odot}/\tau_{A})$  on  $m_{\chi}$ and $m_{\phi}$
for $\varepsilon_{\gamma}=10^{-9}$ (left) and $10^{-10}$ (right).
The regions circled by thick lines are the parameter ranges in
which the DM number $N_{\chi,\bar{\chi}}$ have not yet reached the equilibrium in the current
epoch. 
The region to the left of the dash line is the parameter space that DM evaporation takes place. This is the
non-detection region for the indirect DM search.
The upper panel is for the isospin symmetric case while the lower one
is for the isospin violation case.}
\end{figure}

\begin{figure}
\centering{}\includegraphics[width=0.43\textwidth]{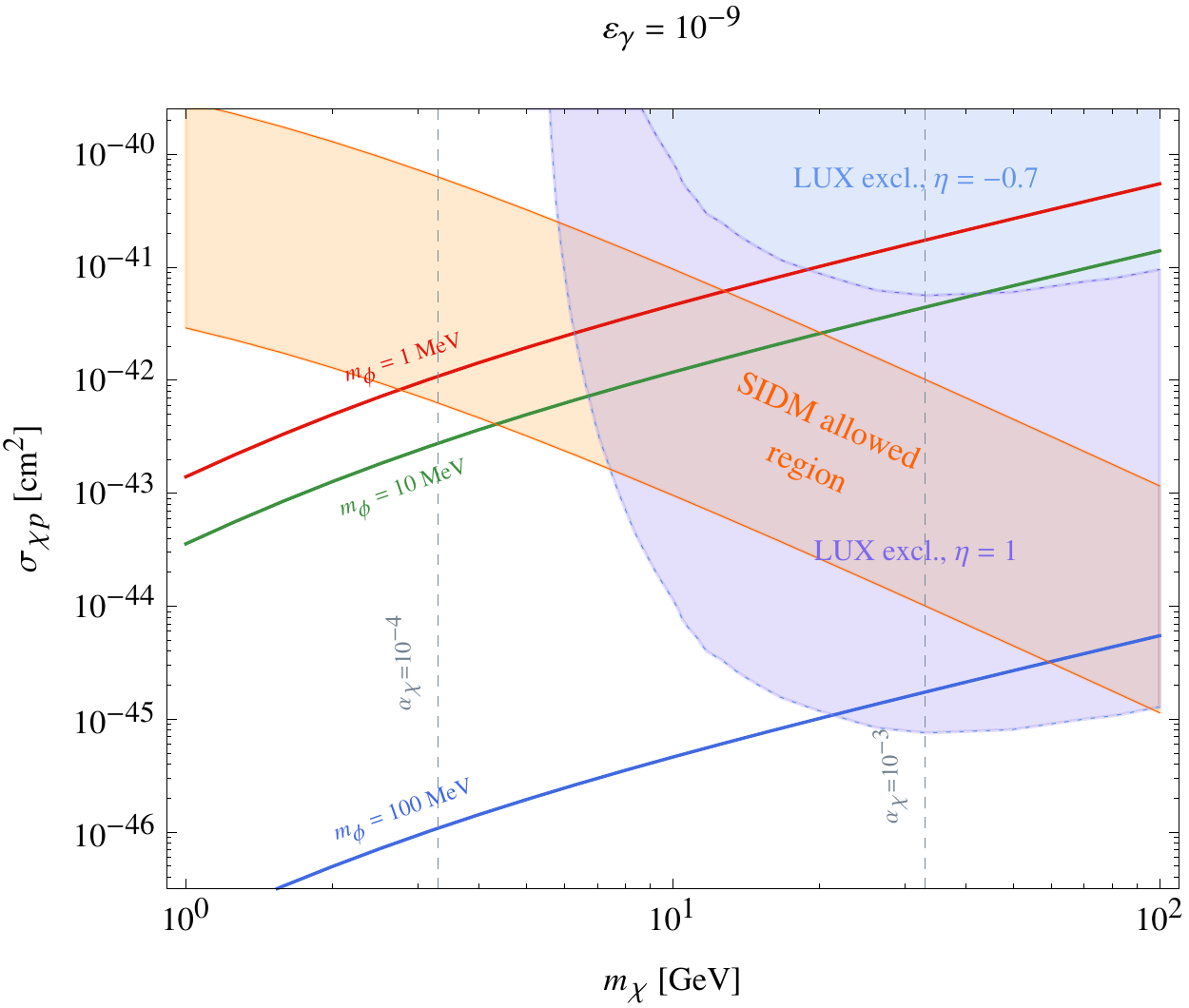}\quad{}\includegraphics[width=0.43\textwidth]{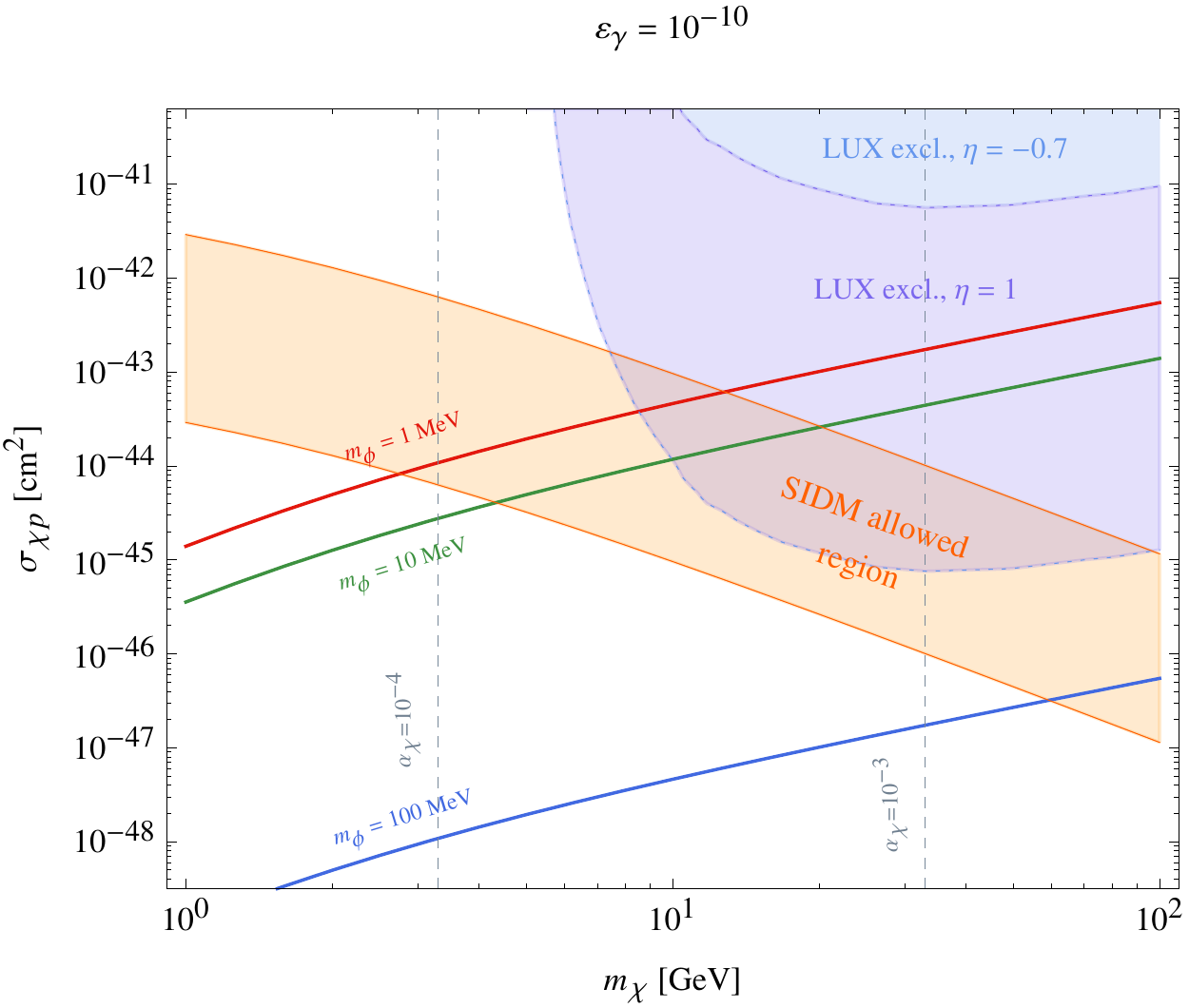}\protect\caption{\label{fig:chi-p_pred}The theoretical predictions of $\sigma_{\chi p}$
with $m_{\phi}=1\,{\rm MeV}$ (red), $10\,{\rm MeV}$ (green) and
$100\,{\rm MeV}$ (blue) are given in solid lines with $\varepsilon_{\gamma}=10^{-9}$
(left panel) and $10^{-10}$ (right panel). The light and deep purple regions
are excluded by LUX \cite{Akerib:2013tjd} for isospin symmetric and isospin violating scenarios, respectively. The orange band is the SIDM allowed region \cite{Vogelsberger:2012ku,Zavala:2012us,Rocha:2012jg,Peter:2012jh}. See the main text for details.}
\end{figure}

To illustrate the role of the light force carrier $m_{\phi}$, 
we first identify the parameter range on the $m_{\chi}-m_{\phi}$ plane in which $N_{\chi}(t)$ has already reached to the equilibrium at the present epoch. In the hidden $U(1)$ model, isospin violation is generally
introduced since $\eta$ can take any value~\cite{Frandsen:2011cg}. 
However, to simplify
our discussions, we consider two extreme scenarios in our calculation. One is the isospin
symmetric scenario with $\eta=1$, and the other corresponds to $\eta=-0.7$ which minimizes DM-Xenon
cross section for a fixed $\sigma_{\chi p}$~\cite{Feng:2013vod}. 
The first scenario corresponds to $\varepsilon_{\gamma}/\varepsilon_{Z}=-0.65$ while the 
second one corresponds to $\varepsilon_{\gamma}/\varepsilon_{Z}=5.65$.  

In Fig.~\ref{fig:eq_time}, we show the equilibrium regions for $\varepsilon_{\gamma}=10^{-9}$
and $10^{-10}$ \cite{Bjorken:2009mm,Lin:2011gj} with $\eta=1$ and $\eta=-0.7$, respectively. The dense color regions
circled by the thick lines are the parameter ranges where the equilibrium has not yet reached
in the current epoch. We quantify such regions by $\tanh(t/\tau_{A})<0.99$. To the left of the gray dashed line, DMs in the Sun evaporate 
and consequently unable to produce detectable signals.
In contrast, the equilibrium is between the capture and the annihilation to the right of the gray dashed line. In this region, the captured DM number is at the maximum, hence the annihilation rate is at its maximum as well.
It is of interest to compare the size of non-equilibrium region for different combinations of $\varepsilon_\gamma$ and $\eta$ in Fig.~\ref{fig:eq_time}.
One can see from Eq.~(\ref{eq:eq_time}) that a smaller $C_c$ leads to a larger equilibrium time scale $\tau_A$ provided all the other coefficients
are held fixed. Comparing the left and right panels of Fig.~\ref{fig:eq_time}, it is seen that the non-equilibrium region increases tremendously
as $\varepsilon_\gamma$ lowering from $10^{-9}$ to $10^{-10}$. 
In fact, one can see from Eq.~(\ref{eq:sig_XP}) that $\sigma_{\chi p}$ 
decreases by two orders of magnitude when $\varepsilon_\gamma$ decreases by an order of magnitude. 
Hence the DM-nucleus cross section $\sigma_{\chi A}$ and consequently the $C_c$ are also significantly reduced. 
This leads to a larger equilibrium time scale as just argued. For $\varepsilon_{\gamma}=10^{-9}$, the non-equilibrium region also increases
with $\eta$ lowering from $1$ to $-0.7$. This is also due to the suppression of $\sigma_{\chi A}$ caused by isospin violation. 
However, with $\varepsilon_{\gamma}=10^{-10}$, 
the non-equilibrium region does not change noticeably when $\eta$ changes from $1$ to $-0.7$. For such an 
$\varepsilon_{\gamma}$, one can show that it is $C_s$ rather than $C_c$ that dictates the equilibrium time scale $\tau_A$., i.e., the DM self-interaction
dominates the capture process. 

The theoretical predictions on $\sigma_{\chi p}$ are shown in Fig.~\ref{fig:chi-p_pred} with colored solid lines. These predictions 
are according to Eq.~(\ref{eq:sig_XP}) and the additional momentum transfer suppression given by Eq.~(\ref{eq:sigma_q}). It is seen that $\sigma_{\chi p}$ depends on $\varepsilon_\gamma$, $\alpha_{\chi}$ and 
$m_{\phi}$. However, $\alpha_{\chi}$ is related  to $m_{\chi}$ by $\alpha_{\chi}\approx3.3\times10^{-5}~(m_{\chi}/{\rm GeV})$ from the thermal relic density of symmetric DM \cite{Kaplinghat:2013yxa,Feng:2008ya}. 
The regions excluded by LUX \cite{Akerib:2013tjd} for the scenarios of isospin symmetry and isospin violation
are displayed for comparisons. The orange band is the allowed region of the SIDM parameter space \cite{Vogelsberger:2012ku,Zavala:2012us,Rocha:2012jg,Peter:2012jh}.

The value of $\eta$ affects the DM-nucleus cross section as seen from Eq.~(\ref{eq:sig_XA}). Hence
$\eta\neq1$ (isospin violation) may weaken the direct search bound on $\sigma_{\chi p}$ as well as suppress the
capture rate for both the Sun and the Earth for a fixed $\sigma_{\chi p}$ as discussed in Refs.~\cite{Feng:2011vu,Gao:2011bq,Lin:2014hla}.
It is clearly seen that the LUX bound on $\sigma_{\chi p}$ is significantly weaken for $\eta=-0.7$.

\begin{figure}
\begin{centering}
\includegraphics[width=0.43\textwidth]{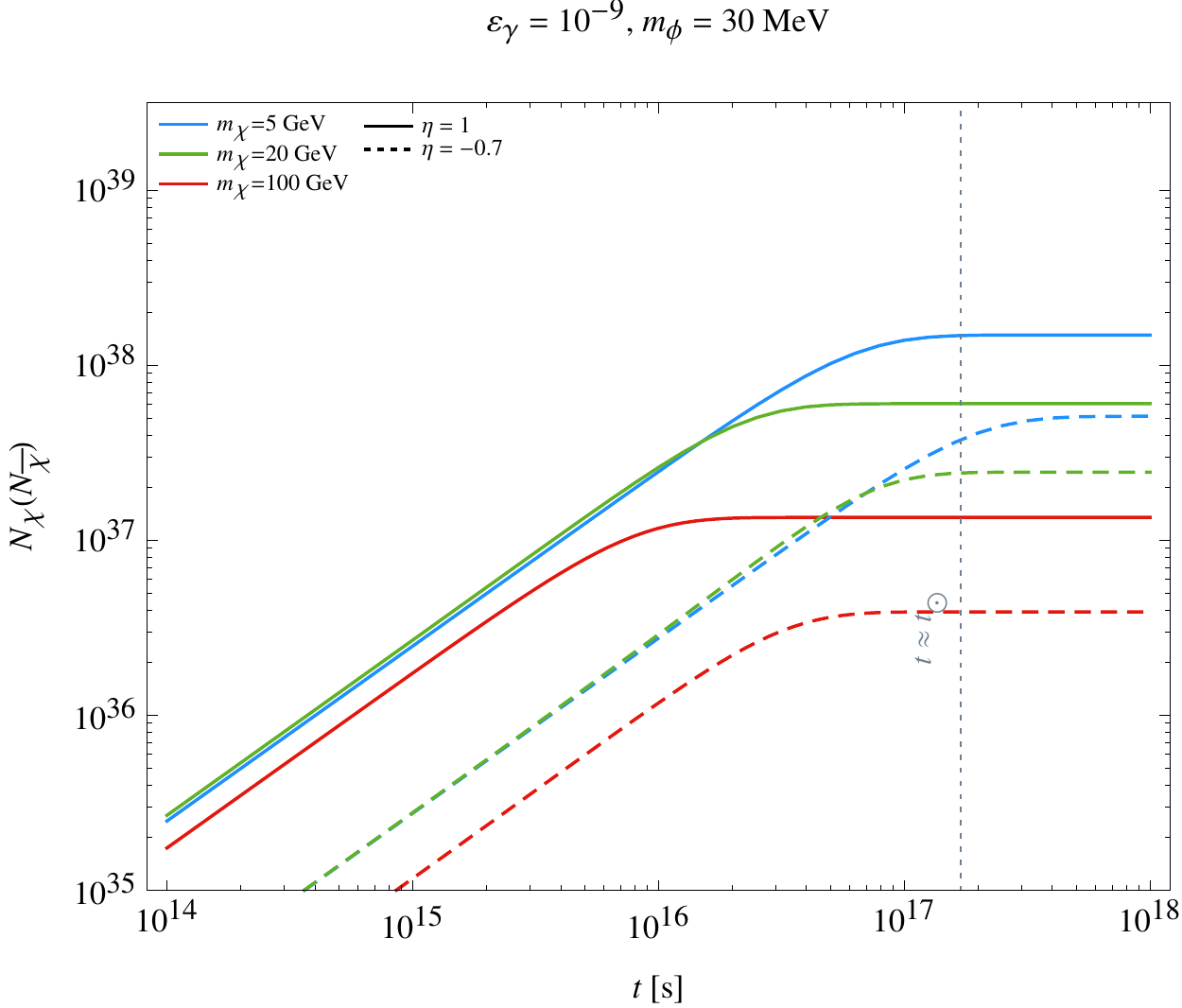}\quad{}\includegraphics[width=0.43\textwidth]{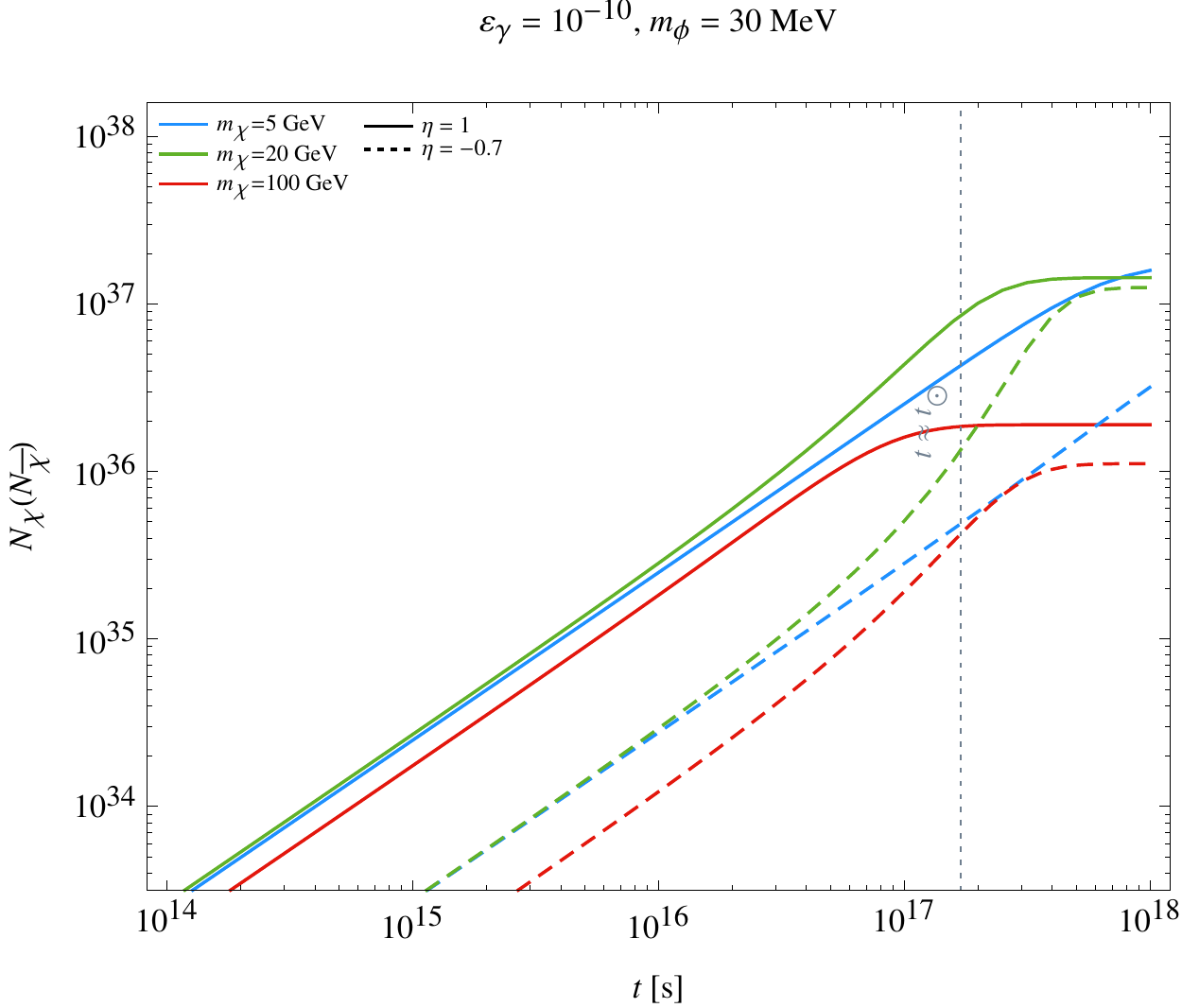}
\par\end{centering}
\begin{centering}
\includegraphics[width=0.43\textwidth]{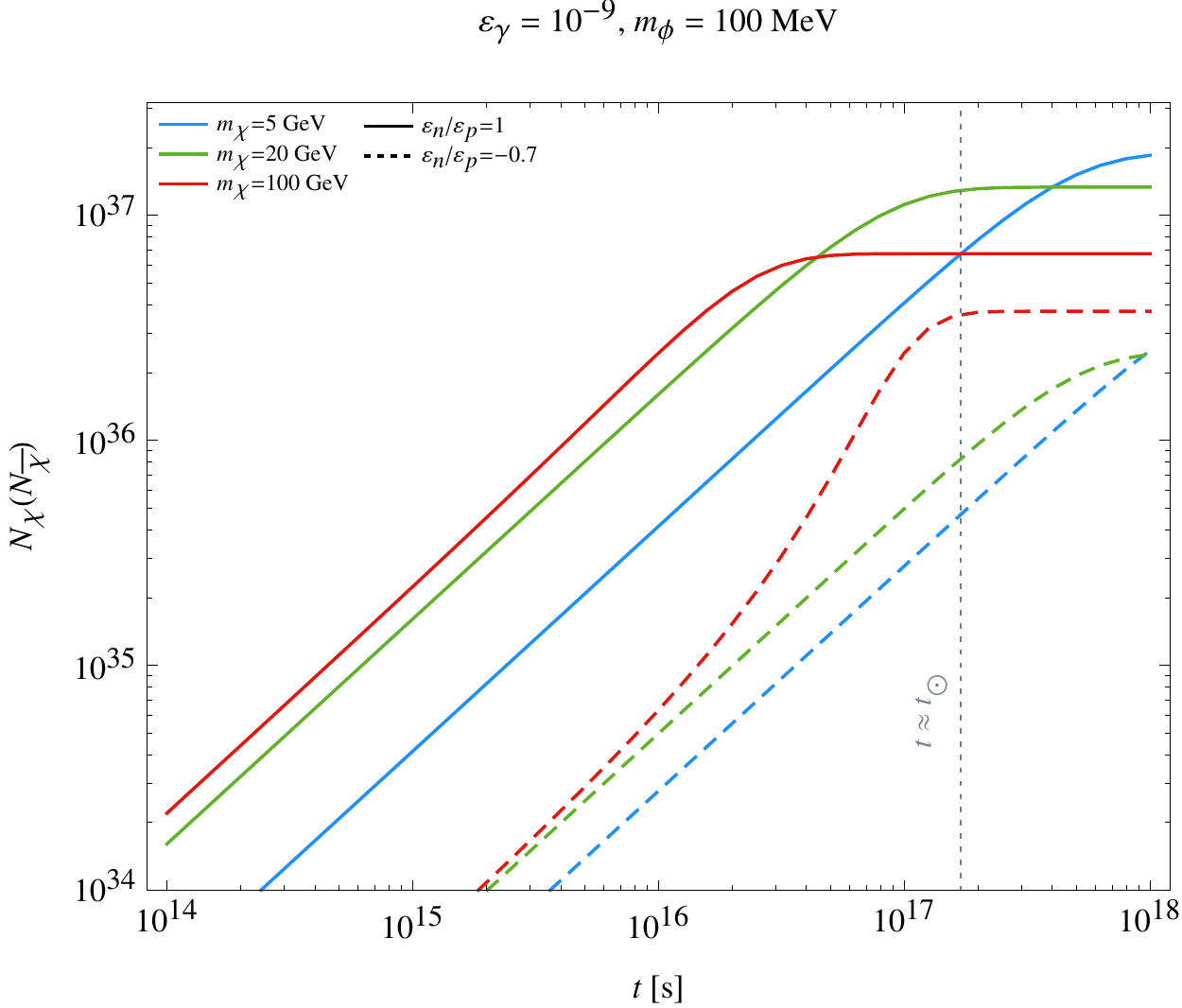}\quad{}\includegraphics[width=0.43\textwidth]{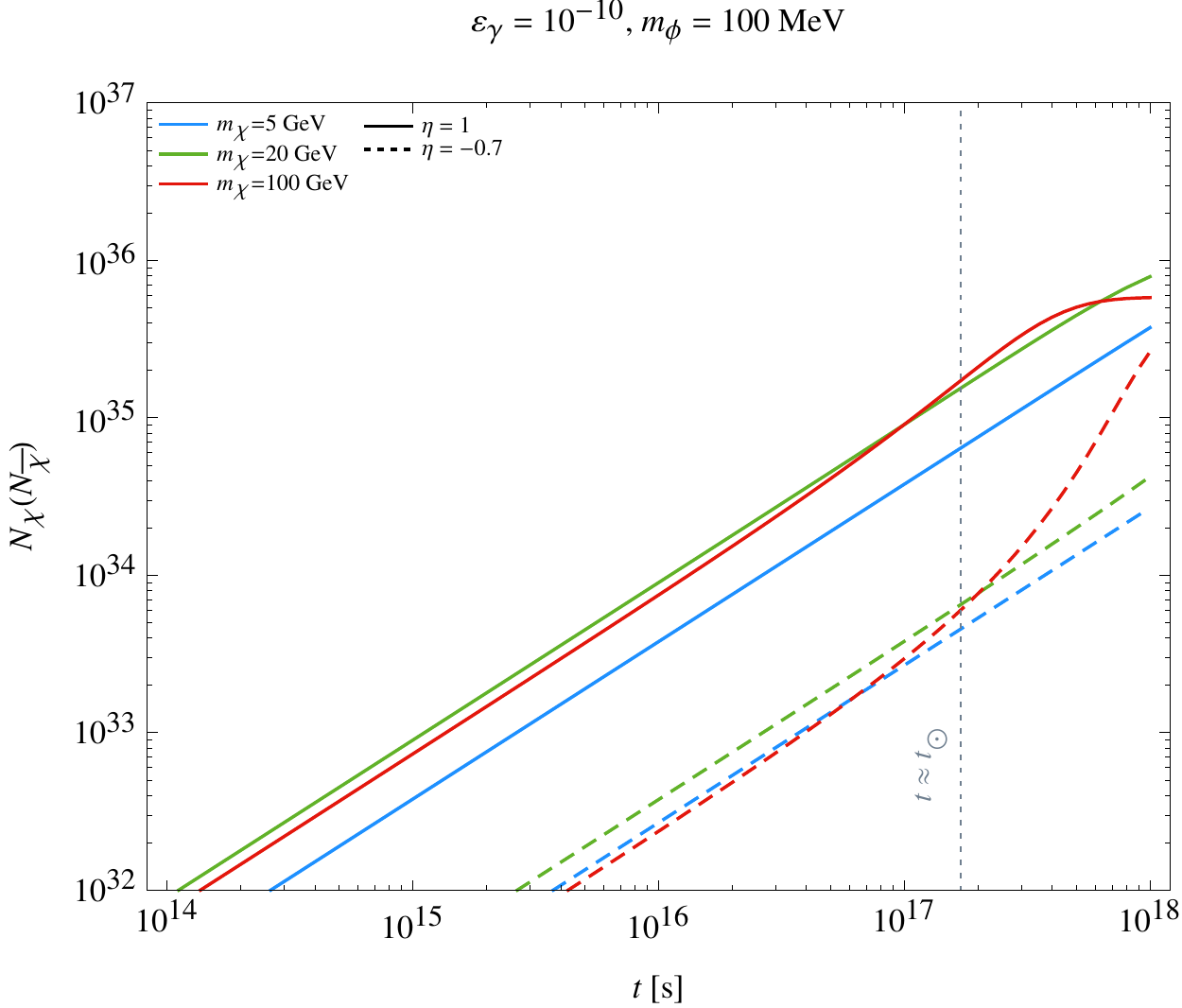}
\par\end{centering}
\protect\caption{\label{fig:N_chi}The $N_{\chi}(N_{\bar{\chi}})$ for $\varepsilon_{\gamma}=10^{-9}$
(left column) and $10^{-10}$ (right column). The upper panels are
for $m_{\phi}=30\,{\rm MeV}$ and the lower panels are for $m_{\phi}=100\,{\rm MeV}$.
 Solid
lines are for isospin symmetric case  and dashed lines are for isospin violated case. Blue, green and
red curves are for $m{}_{\chi}=5\,{\rm GeV},\,20\,{\rm GeV}\,{\rm and}\,100\,{\rm GeV}$,
respectively. The solar age ($t_{\odot}$) is indicated by the gray
dashed line.}
\end{figure}

The evolution behaviors of $N_{\chi}$ with
different $m_{\chi},~m_{\phi}$ and $\eta$
are shown in Fig.~\ref{fig:N_chi}. On the upper left
panel, the difference in $N_{\chi}$ between $\eta=1$ and $\eta=-0.7$ can be understood by
the following arguments. First, the coefficient $C_c$ with $\eta=-0.7$ is suppressed compared to $C_c$ with $\eta=1$.  Hence $N_{\chi}$ with $\eta=1$ is greater than $N_{\chi}$ with $\eta=-0.7$ no matter how large $C_s$ is. However, as $m_{\chi}$ increases, $C_s$ becomes more and more dominant over $C_c$ since $\sigma_{\chi\chi}$ in $C_c$ grows as 
$m_{\chi}^4$ while $\sigma_{\chi A}$ in $C_c$ grows as $m_{\chi}$ (this follows from Eqs.~(\ref{eq:sig_X-barX}) and (\ref{eq:sig_XA})  and the fact that $\alpha_{\chi}$ grows as $m_{\chi}$). Hence the reduction of $C_c$ by $\eta=-0.7$ does not produce significant effect on $N_{\chi}$, given a 
very dominant $C_s$. This explains why $N_{\chi}$ is less sensitive to $\eta$ for larger $m_{\chi}$.  
Comparing upper right panel with the upper left one, we find $N_{\chi}$ is less sensitive to $\eta$ in the former case for the same $m_{\chi}$. 
The DM-nucleus cross section $\sigma_{\chi A}$ for the upper right panel is suppressed due to a smaller $\varepsilon_{\gamma}$. 
Hence the effect from $\eta$ becomes less significant. 
For lower left and lower right panels of  Fig.~\ref{fig:N_chi}, we take $m_{\phi}=100\,{\rm MeV}$. 
In this case, both $\sigma_{\chi A}$ and $\sigma_{\chi\chi}$ are suppressed. Hence the number of captured DM is much less and 
the equilibrium time scale is much larger. 

\section{Testing SIDM model in IceCube-PINGU}

\subsection{Neutrino signal and the atmospheric background}

Neutrino signals arise from the decays of $\phi$ which is produced by DM annihilation, i.e., $\chi\bar{\chi}\rightarrow\phi\phi\rightarrow4\nu$. DM annihilate to $ZZ$ and $\gamma\gamma$ modes are suppressed by small couplings between $\chi$-Z and $\chi$-$\gamma$ respectively. 
For the hidden $U(1)$ gauge model considered in this paper, $\phi$ decays into SM particles via $\varepsilon_{\gamma}$ and $\varepsilon_{Z}$ mixings. Due to the kinematics constraint, $e^+e^-$ and neutrinos are the only decay products of $\phi$ and the corresponding decay widths through photon and $Z$ mixings are given in Ref.~\cite{Kaplinghat:2013yxa}. 
The branching ratio for $\phi\to \nu\bar{\nu}$ is determined by the 
relative magnitudes of mixing parameters $\varepsilon_{\gamma}$ and $\varepsilon_{Z}$. Therefore it is determined by the parameter $\eta$ through Eqs. ~(\ref{eq:e_p})
and (\ref{eq:e_p}). We find  BR($\phi \rightarrow \nu\bar{\nu}$) $\approx$ $75\%, 39\%, 48\%$, and $67\%$ for 
 $\eta = 1, -0.3, -0.5$, and $-0.7$, respectively. Furthermore, although $\phi$ is produced on-shell 
in the solar center, it does not decay instantly but instead propagates for a certain distance. 
The lifetime of $\phi$ is constrained by BBN, $\tau_{\phi}\lesssim\mathcal{O}(1)\,{\rm s}$~\cite{Kaplinghat:2013yxa,Lin:2011gj}. Considering a $30$ MeV $\phi$ with $5$ GeV 
energy (corresponding to 
$m_{\chi}=10$ GeV), the decay time of $\phi$ would be less than $170$ s. One can easily estimated that, for a $\phi$ with $170$ s of decay time,
it travels for roughly $5\times 10^7$ km before decaying into the electron-position or the neutrino pair. The decay point is already outside the Sun but not yet reaching to the Earth. 
Given the distance between the Sun and the Earth at $1.5\times 10^8$ km, those $\phi$ with $30$ MeV of mass and moving toward the Earth shall decay between Sun and Earth provided $E_{\phi}$ is less than $150$ GeV. Hence the neutrino flux will be observed by the terrestrial detector. We have so far made our argument with $\tau_{\phi}=1$ s. For 
$\tau_{\phi}\ll 1$ s,  the decay point of $\phi$ could be inside the Sun. In this case, the neutrino propagating distance from source to 
terrestrial detector becomes larger. However this does not affect the oscillation of neutrinos because neutrino propagating distances in both cases are 
much larger than the neutrino oscillation length.  Hence our results on neutrino event rates are not affected. 

The argument in the last paragraph  
assumes $\phi$ propagating freely inside the Sun. To justify this assumption, we study the interaction between $\phi$ and hydrogen, which occurs through
the mixing between $\phi$ and $\gamma$ and the subsequent $\gamma p$ scattering. The total $\gamma p$ scattering cross section at $\sqrt{s}=\sqrt{10}$ GeV is
about $0.1$ mb~\cite{Baldini:1988ti}. Taking $\varepsilon_{\gamma}=10^{-9}$, we have $\phi p$ total cross section as small as $10^{-46}$ cm$^2$. 
With the average hydrogen number density in the Sun about $6\times 10^{23}$/cm$^3$, the mean free path of $\phi$ is much greater than the radius
of the Sun. We note that the consideration of other chemical elements in the Sun should shorten the 
mean free path of $\phi$ slightly. Nonetheless, this distance scale remains  much greater than the radius
of the Sun. 

Since each $\phi$ stays at the same direction before its decay, one can view those neutrinos produced by $\phi$ decays as being originated from the 
core of the Sun. Hence    
we can write the neutrino flux as
\begin{equation}
\frac{d\Phi_{\nu_{i}}}{dE_{\nu_{i}}}=P_{\nu_{j}\rightarrow\nu_{i}}(E_{\nu})\frac{\Gamma_{A}}{4\pi R^{2}}\frac{dN_{\nu_{j}}}{dE_{\nu_{j}}} ,
\end{equation}
with $P_{\nu_{j}\rightarrow\nu_{i}}(E_{\nu})$ the neutrino oscillation
during the propagation, $\Gamma_A$ the DM annihilation rate, $R$ the distance between Sun and Earth, and $dN_{\nu_{j}}/dE_{\nu_{j}}$
the neutrino spectrum from each annihilation. The energy distribution $dN_{\nu_{j}}/dE_{\nu_{j}}$ for neutrinos produced by $\phi$ decays is not a simple Dirac-$\delta$ function,
since $\phi$ is highly boosted. It has been shown that \cite{Dasgupta:2012bd}
	\begin{equation}\label{eq:dNdE}
	\frac{dN_{\nu}}{dE_{\nu}}=\frac{4}{\Delta E}\Theta (E_\nu-E_-)\Theta (E_+-E_\nu),
	\end{equation}
where $\Theta$ denotes the Heaviside step function. $E_{\pm}=(m_\chi \pm \sqrt{m_\chi^2-m_\phi^2})/2$ are the maximum and minimum of the neutrino energy. $\Delta E\equiv \sqrt{m_\chi^2-m_\phi^2}$ denotes the width of the energy spectrum.

The neutrino event rate in the detector is given by
\begin{equation}
N_{\nu}=\int_{E_{{\rm th}}}^{m_{\chi}}\frac{d\Phi_{\nu_{i}}}{dE_{\nu_{i}}}A_{\nu}(E_{\nu})dE_{\nu}d\Omega~,\label{eq:event_rate}
\end{equation}
where $E_{{\rm th}}$ is the detector threshold energy, $A_{\nu}(E_{\nu})$
the detector effective area and $\Omega$ the solid angle. We study
both muon track events and cascade events induced by neutrinos. The
DeepCore detector extends the IceCube capability to probe $E_{\nu}$ down to $10\,{\rm GeV}$
\cite{Collaboration:2011ym} and the future PINGU detector will further lower down the energy threshold $E_{\rm th}$ to $\mathcal{O}(1)\,{\rm GeV}$~\cite{Aartsen:2014oha}. The angular resolution for IceCube-PINGU detector at $E_{\nu}=5$ GeV  is roughly $10^{\circ}$. Hence we consider neutrino events arriving from the solid angle 
range $\Delta{\Omega}=2\pi(1-\cos\psi)$ surrounding the Sun with $\psi=10^{\circ}$.
The detector effective area of IceCube is expressed as
\begin{equation}
A_{\nu}(E_{\nu})=V_{{\rm eff}}\frac{N_{A}}{M_{{\rm ice}}}[n_{p}\sigma_{\nu p}(E_{\nu})+n_{n}\sigma_{\nu n}(E_{\nu})] 
\end{equation}
with $V_{\rm eff}$ the detector effective volume \cite{Collaboration:2011ym,Aartsen:2014oha}, $N_{A}$ the Avogadro number, $M_{{\rm ice}}$ the molar
mass of ice, $n_{p,n}$ the number density of proton/neutron per mole
of ice, and $\sigma_{\nu p,n}$  the neutrino-proton/neutron cross
section. One simply makes the replacement $\nu\rightarrow\bar{\nu}$ for
anti-neutrino.

The atmospheric background is similar to Eq.~(\ref{eq:event_rate}),
by replacing $d\Phi_{\nu}/dE_{\nu}$ with atmospheric neutrino flux,
\[
N_{{\rm atm}}=\int_{E_{{\rm th}}}^{E_{{\rm max}}}\frac{d\Phi_{\nu}^{{\rm atm}}}{dE_{\nu}}A_{\nu}(E_{\nu})dE_{\nu}d\Omega.
\]
The $d\Phi_{\nu}^{{\rm atm}}/dE_{\nu}$ is taken from Ref.~\cite{Honda:2006qj}. We set
$E_{{\rm max}}=m_{\chi}$ in order to compare with DM signal.

\subsection{The IceCube sensitivity to the mass of the force carrier}

\begin{figure}
	\begin{centering}
		\includegraphics[width=0.43\textwidth]{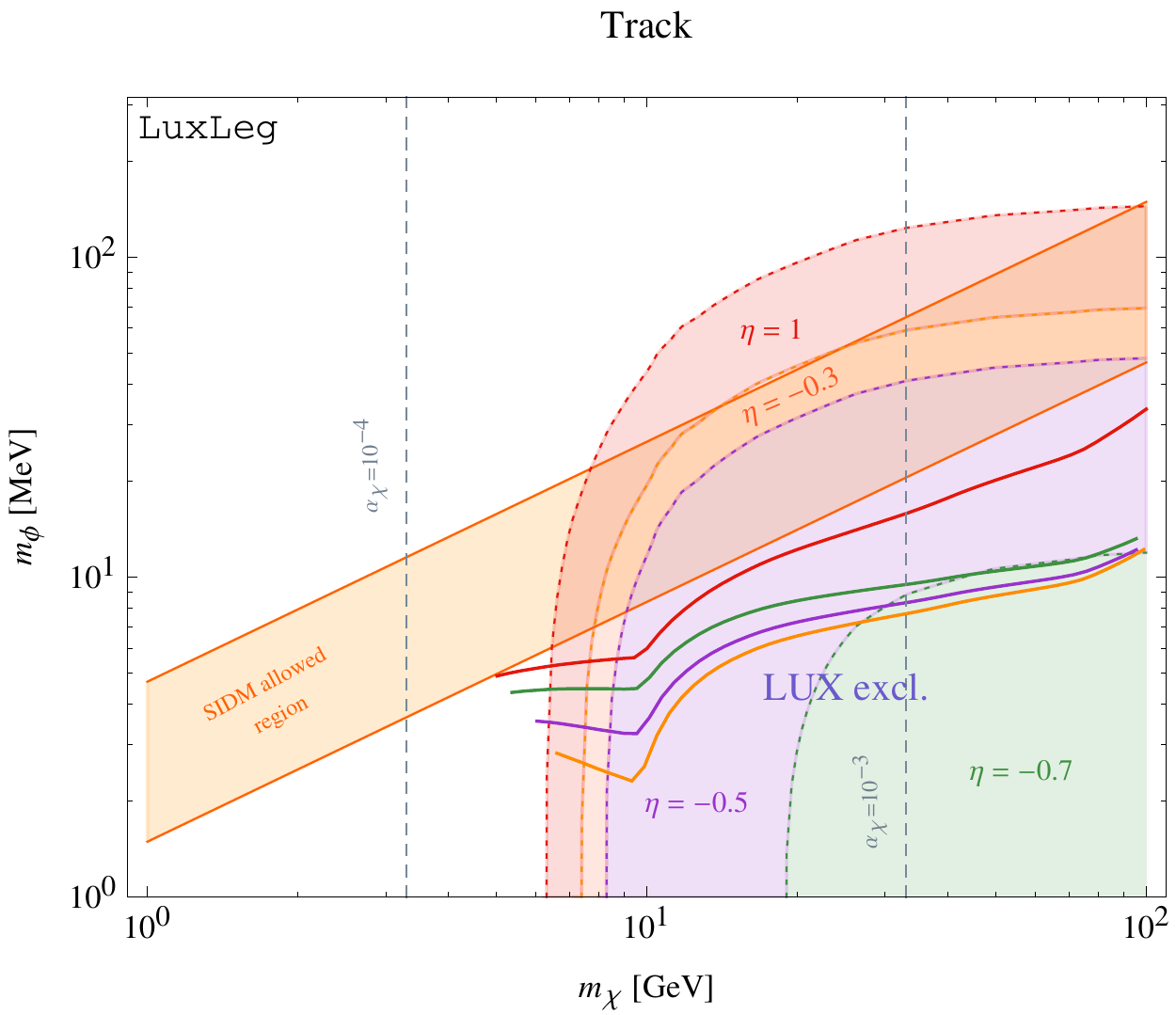}\quad{}\includegraphics[width=0.43\textwidth]{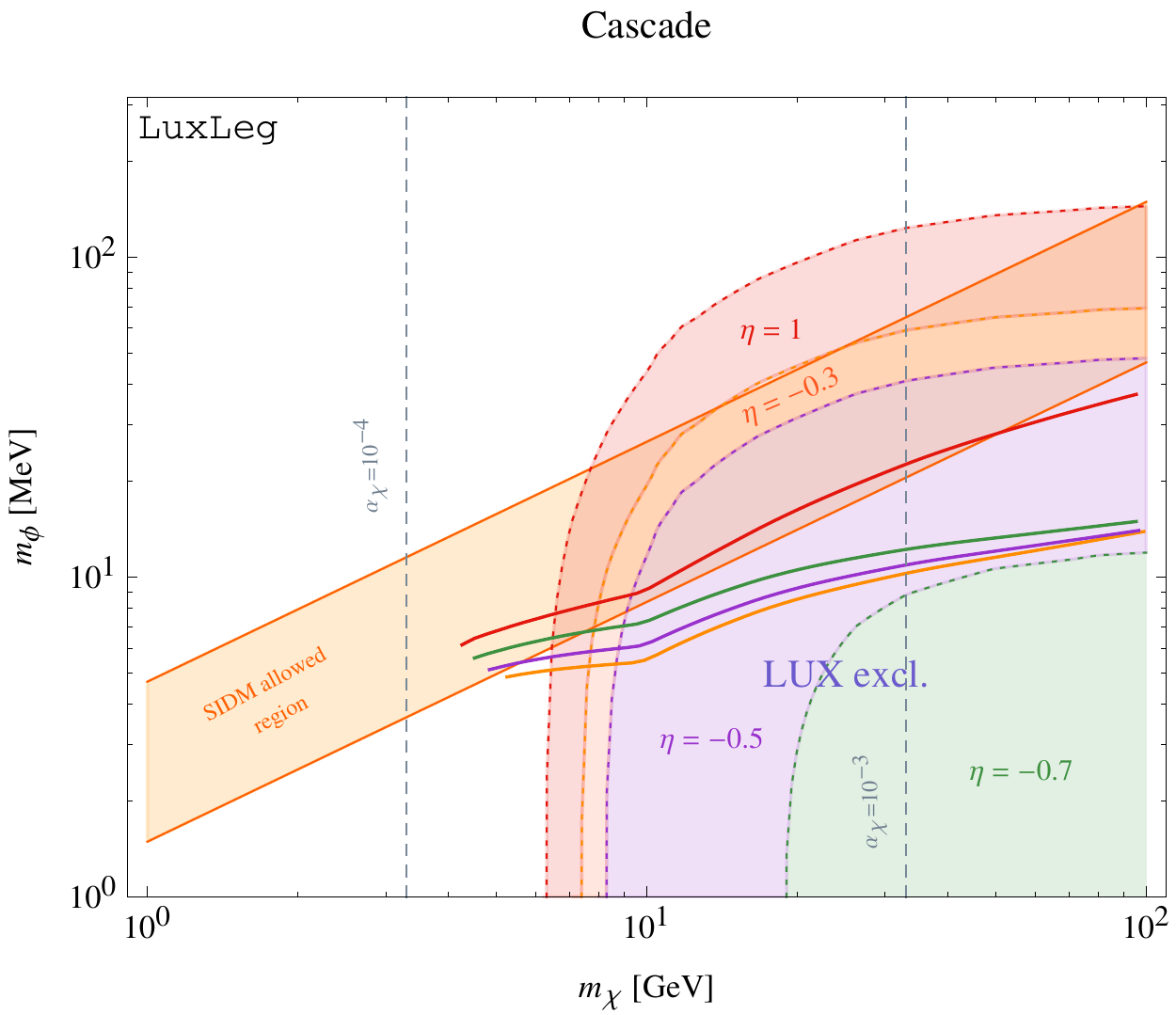}
		\par	\end{centering}
	\protect\caption{\label{fig:m_phi_sensitivity}
		The IceCube-PINGU sensitivities to $m_{\phi}$ for different values of $\eta$. The left panel is for track events and the right panel is for cascade events. The sensitivity curves and LUX excluded regions correspond to $\varepsilon_\gamma =10^{-9}$. The results corresponding to $\varepsilon_\gamma =10^{-10}$ are not shown since BBN constraint rules out most of the $m_\phi$ parameter space plotted here.
	}
\end{figure}

\begin{table}
	\begin{centering}
		\begin{tabular}{ccccc}
			\hline 
			& \multicolumn{2}{c}{Track} & \multicolumn{2}{c}{Cascade}\tabularnewline
			\hline 
			$E_{{\rm max}}$ {[}GeV{]} & $N_{\nu}^{{\rm atm}}$ & $N_{\nu}^{{\rm DM}}$ & $N_{\nu}^{{\rm atm}}$ & $N_{\nu}^{{\rm DM}}$\tabularnewline
			\hline 
			\hline 
			5 & 7146 & 76 & 9874 & 90\tabularnewline
			10 & 10280 & 91 & 13775 & 105\tabularnewline
			50 & 21680 & 132 & 21803 & 132\tabularnewline
			70 & 23584 & 138 & 23111 & 136\tabularnewline
			100 & 26610 & 146 & 24363 & 140\tabularnewline
			\hline 
		\end{tabular}
		\par\end{centering}
	
	\caption{\label{tab:signal&bkg}The annual signal and background event numbers for reaching 
	$2\sigma$ detection significance in 5 years.   }
\end{table}

We present the sensitivity as a $2\sigma$ detection significance
in 5 years, calculated with the formula,
\[
\frac{s}{\sqrt{s+b}}=2.0 ,
\]
where $s$ is the DM signal, $b$ is the atmospheric background, and
2.0 is referring to the $2\sigma$ detection significance. The detector
threshold energy $E_{{\rm th}}$ is set to be 1 GeV, which will be achieved
by the future PINGU detector. However, for $E_{\nu}>10\,{\rm GeV}$,
we adopt the DeepCore effective area in our calculation. The annual signal and background event numbers for reaching $2\sigma$ detection significance in 5 years are listed in Table~\ref{tab:signal&bkg}. We present the track and cascade event numbers separately.

In Fig.~\ref{fig:m_phi_sensitivity}, we present the IceCube-PINGU sensitivities to $m_\phi$ with different values of $\eta$ in colored solid lines. Regions above these lines are beyond the reach of the detector within 5 years for a $2\sigma$ detection significance. Furthermore, each sensitivity curve terminates at the evaporation mass scale
below which the DM signature from the Sun is suppressed. The left panel is for track events and the right one is for cascade events. BBN excludes $m_\phi<20~\rm MeV$ for $\varepsilon_\gamma=10^{-10}$, while it excludes $m_\phi<0.3~\rm MeV$ for $\varepsilon_\gamma=10^{-9}$. For simplicity, we only present results for $\varepsilon_\gamma=10^{-9}$  since BBN constraint for $\varepsilon_\gamma=10^{-10}$ rules out most of the $m_{\phi}$ parameter space shown here. The orange band is SIDM allowed region.
Those color shaded regions are excluded by LUX for different $\eta$ values. The gray dashed lines indicate the relation between $\alpha_{\chi}$ and $m_{\chi}$ required by the thermal relic density.


It is of interest to compare results with different $\eta$ values. It is seen that the sensitivity curve 
with $\eta=1$ differs significantly from those with other $\eta$ values.  However, sensitivity curves  with  $-0.3\leq\eta\leq-0.7$ do not differ much from one another. Clearly $C_s$ rather than $C_c$ dominates the DM capture in this range of $\eta$. 

Comparing left and right panels of Fig.~\ref{fig:m_phi_sensitivity}, one can see that cascade events can probe a larger part of
 $m_{\chi}-m_{\phi}$ parameter space. Furthermore, only cascade events can probe into the SIDM allowed region. Comparing IceCube-PINGU sensitivity with
 LUX constraint, the former complements the latter for low-mass DM until the evaporation mass scale $m_\chi \lesssim 4~\rm GeV$.
Such a complementarity also occurs due to isospin violation. It is well known that the direct search bound on $\sigma_{\chi p}$ could be significantly weaken by isospin violation. In fact, one can see that the LUX excluded region shrinks as $\eta$ decreases from $1$ to $-0.7$. On the other hand, the indirect DM signature from the Sun is less affected by $\eta$. Hence the complementarity
of two searches becomes more apparent when $\sigma_{\chi p}$ is subject to more severe isospin violation 
suppression. 

\section{Summary}

We have proposed to test SIDM model with hidden $U(1)$ gauge symmetry 
by searching for DM-induced neutrino signature from the Sun. The SIDM model under consideration gives rise to
DM-nucleus scattering, DM-DM elastic scatterings, and DM-DM annihilation. These three processes 
determine the neutrino flux resulting from DM annihilation in the Sun.  We have presented 
the IceCube-PINGU sensitivities to the parameter space of SIDM model. We compare these sensitivities to
existing constraints set by LUX experiment. We have shown that the indirect search complements the direct one 
in two ways. First, the indirect search is more sensitive to light DM in GeV mass range. Second, the direct search constraint on SIDM parameter space can be significantly weaken by isospin violation while
the indirect search sensitivity is less affected by this effect.    


\section*{Acknowledgments\label{sec:ackn}}
 We thank R. Laha for pointing out Ref.~\cite{Dasgupta:2012bd} to us. CSC is supported by the National
Center for Theoretical Sciences, Taiwan; GLL and YHL are supported by Ministry of Science and Technology, Taiwan under 
Grant No. 103-2112-M-009 -018.

\end{document}